\documentclass[pdflatex,sn-mathphys-num]{sn-jnl}

\usepackage{graphicx}%
\usepackage{multirow}%
\usepackage{amsmath,amssymb,amsfonts}%
\usepackage{amsthm}%
\usepackage{mathrsfs}%
\usepackage[title]{appendix}%
\usepackage{xcolor}%
\usepackage{textcomp}%
\usepackage{manyfoot}%
\usepackage{booktabs}%
\usepackage{algorithm}%
\usepackage{algorithmicx}%
\usepackage{algpseudocode}%
\usepackage{listings}%
\usepackage{braket}
\usepackage[justification=centering]{caption}

\theoremstyle{thmstyleone}%
\theoremstyle{thmstyletwo}%

\theoremstyle{thmstylethree}%
\begin{document}

\title[Article Title]{Hybrid Quantum-Classical Selective State Space Artificial Intelligence }

\author[1]{\fnm{Amin} \sur{Ebrahimi}}

\author*[1]{\fnm{Farzan} \sur{Haddadi}}\email{farzanhaddadi@iust.ac.ir}

\affil[1]{\orgdiv{School of Electrical Engineering}, \orgname{Iran University of Science \& Technology},\orgaddress{\city{Tehran}, \country{Iran}}}

\abstract{Hybrid Quantum–Classical (HQC) algorithms constitute one of the most effective paradigms for exploiting the computational advantages of quantum systems in large-scale numerical tasks. By operating in high-dimensional Hilbert spaces, quantum circuits enable exponential speed-ups and provide access to richer representations of cost landscapes compared to purely classical methods. These capabilities are particularly relevant for machine learning, where state-of-the-art models especially in Natural Language Processing (NLP) suffer from prohibitive time complexity due to massive matrix multiplications and high-dimensional optimization.
	
In this manuscript, we propose a Hybrid Quantum–Classical selection mechanism for the Mamba architecture, designed specifically for temporal sequence classification problems. Our approach leverages Variational Quantum Circuits (VQCs) as quantum gating modules that both enhance feature extraction and improve suppression of irrelevant information. This integration directly addresses the computational bottlenecks of deep learning architectures by exploiting quantum resources for more efficient representation learning.
	
We analyze how introducing quantum subroutines into large language models (LLMs) impacts their generalization capability, expressivity, and parameter efficiency. The results highlight the potential of quantum-enhanced gating mechanisms as a path toward scalable, resource-efficient NLP models, in a limited simulation step. Within the first four epochs on a reshaped MNIST dataset with input format (batch, 784, d\_model), our hybrid model achieved 24.6\% accuracy while using one quantum layer and achieve higher expressivity, compared to 21.6\% obtained by a purely classical selection mechanism.
we state No founding }


\maketitle

\section{Introduction}\label{Intro}
Hybrid Quantum–Classical (HQC) algorithms represent a new computational paradigm for the near-term era of quantum computing \cite{preskill2018}.
By combining the strengths of quantum and classical resources, HQCs provide a flexible architecture where quantum modules contribute superior properties—such as access to richer Hilbert spaces \cite{schuld2019quantum}, superposition of states, entanglement \cite{nielsen2010quantum}, and quantum channel transformations \cite{beer2020training}—while classical modules manage large-scale processing, memory storage, and optimization with lower overhead.

This hybrid structure has been applied across a wide range of domains, including the Quantum Approximate Optimization Algorithm (QAOA) \cite{farhiQuantumApproximateOptimization2014a}, Variational Quantum Eigensolver (VQE) \cite{cao2019}, and Variational Quantum Linear Solvers (VQLS) \cite{HHL2009}, which are central tools for tackling optimization problems. Consequently,  Machine Learning, formulated as an optimization problem, can benefit from quantum algorithms. Notable candidates include quantum linear solvers \cite{bravo2023,childs2017,gilyen2019}, Variational Quantum Circuits \cite{arunachal2017,biamonte2018}, and Quantum Neural Networks (QNNs) \cite{schuld2018c,beer2020training,havlicek2019,abbas2021,huang2022}.

In particular, VQCs \cite{alam2021,benedetti2019,havlicek2019,mitarai2018b,schuld2020} employ trainable parametrized gates in their ansatz, allowing them to act as information loaders and feature manipulators—similar to, but more powerful than, Multilayer Perceptrons (MLPs)—thereby increasing the expressivity and effective dimension of the total ansatz \cite{abbas2021}, often achieving equal or superior performance to classical MLPs with fewer parameters. 
While the exponentially large Hilbert space explored by VQCs offers rich representational capacity that may aid generalization, but somehow could cause serious problem well-known vanishing gradient which is known in quantum literature as Barren-Platueaus effect \cite{mcclean2018}, where gradients vanish in high-dimensional landscapes, posing a major obstacle for scalability.

Most VQC-based approaches to classical data processing are structured around three components: encoding, ansatz, and measurement.
The encoding layer maps classical data into quantum states, typically through amplitude or angle encoding, though more sophisticated feature maps can be employed to increase structure and density. These encoding choices strongly influence the design of the subsequent ansatz and measurement strategy.  The ansatz itself is usually selected to maximize expressivity and entanglement capability \cite{sim2019expressibility}, while respecting the practical constraints of current quantum hardware, particularly the limitations on qubit connectivity and entanglement depth.

From the perspective of classical sequential learning, the evolution began with Recurrent Neural Networks (RNN), which extracts data from temporal sequences. To address the limitations of vanishing gradients and long-term memory loss, RNNs were extended to Long Short-Term Memory (LSTM) networks \cite{hochreiter1997} and Gated Recurrent Units (GRUs) \cite{cho2014properties}. Hence these gated structures improved sequence modeling, yet challenges persisted when handling very long dependencies in sequences, specially as in genomic or hard correlation long sequences problem.
A major breakthrough came with the introduction of the Transformer architecture \cite{vaswani2017attention}, which surpassed traditional sequential models in accuracy through its attention mechanism. However, the quadratic complexity of pairwise token similarity computations imposed severe computational costs. On the other hand, there was plenty tries to reduce complexity such as Reformer \cite{kitaev2020reformer} which imployed Locality-Sensitive Hashing(LHS), split embedding process and reversible layers. Longformer \cite{beltagy2020longformer} by combined windowed and global sparse attention. Linformer \cite{wang2020linformer}, by reducing key–value dimensionality via learnable linear projections. and Performer \cite{choromanski2020rethinking}, which approximated the softmax attention using positive orthogonal random features. While these alternatives reduced time complexity and improved scalability, they often sacrificed performance by discarding valuable information.
A similar approach to traditional RNNs, emerged through state space modeling.
The S4 model \cite{gu2021efficiently} introduced the HiPPO matrix as parameter initialize strategy and capture long-range dependencies efficiently. Then S5 was proposed \cite{smith2022simplified}, which simplified implementation using diagonal-plus-low-rank techniques. Most recently, Mamba (S6 generation) \cite{gu2023mamba} leveraged Diagonal State Space Models (DSSs) where compacted the initialization core SSM modules also implemented efficient CUDA kernel of selective parallel scan which could even select the processing space GPU-SRAM–VRAM to achieve better training speed, while maintaining lower computational complexity. Mamba integrates the Gated MLP, as a function handler to managing decimation irrelevant data or boost effect critical information, wrapping around simple SSM core (S4) (DSS) and convolution block to improve information flow. 
Although Transformer-based models may still achieve superior performance in certain tasks, state space architectures such as Mamba offer substantial advantages in resource-constrained settings.

Meanwhile, investigating Quantum Deep Learning started when made possible in quantum Devices. \cite{chen2020b} presented a hybrid Quantum-LSTM architecture where used six VQC structure as classical mixer projector gates. Followed by Quantum RNNs, \cite{li2023} reduced circuit depth by reusing qubits across unfolded latent sequences to overcome coherence time. recently \cite{moon2025} improved accuracy  but since the architecture follows traditional RNN it contents with  the vanishing information. however transformer is the best  where quantum computing could be used State space models were considered as an alternative to transformer \cite{basile2017,sipio2021,emmanoulopoulos2022c,comajoancara2024,guo2024quantum,khatri2024} 

In this paper, we introduce HQC instructions for selective state space  model S6 which are very successful in classical domain. This state space model will be an improvement on transformer in lower computational complexity.

\section{Methods}\label{Methods}

\subsection{Classical perspective}
Recurrent Neural Networks structures can extract dependencies (temporal or spatial) between samples in sequences.If we unfold the structure in time,  data $X \in \mathbb{R}^{d}$ is concatenated as  $\{\mathbf{x}_1, \mathbf{x}_2, \dots, \mathbf{x}_T\}$ feed to model as an input and each time step it combines with hidden state using learnable weights $\mathbf{W}_x \in \mathbb{R}^{m \times d}$  which is an affine transformation of input features to hidden state. Translated data, sums with pervious hidden state by another learnable affine transformation $\mathbf{W}_h \in \mathbb{R}^{m \times m}$. in general the relation of each hidden state is:
\begin{equation}
	\mathbf{h}_t = \sigma_h(\mathbf{W}_x \mathbf{x}_t + \mathbf{W}_h \mathbf{h}_{t-1} + \mathbf{b}_h)
	\label{rnn-eq}
\end{equation}

where $\mathbf{h}_t \in \mathbb{R}^m$ is the hidden state and $\sigma_h$ is a nonlinear activation function (e.g., $\tanh$ or $\text{ReLU}$) and $b_h$ is the bias term. By \textit{unfolding in time} for $t = 3$, we obtain:

\[
\begin{aligned}
	\mathbf{h}_1 &= \sigma_h(\mathbf{W}_x \mathbf{x}_1 + \mathbf{W}_h \mathbf{h}_0 + \mathbf{b}_h), \\
	\mathbf{h}_2 &= \sigma_h(\mathbf{W}_x \mathbf{x}_2 + \mathbf{W}_h \mathbf{h}_1 + \mathbf{b}_h), \\
	\mathbf{h}_3 &= \sigma_h(\mathbf{W}_x \mathbf{x}_3 + \mathbf{W}_h \mathbf{h}_2 + \mathbf{b}_h).
\end{aligned}
\]

 hidden state contains compressed information of pervious time step space.It with getting new input, former data is going to vanish. 
A solution to this problem is LSTM \cite{hochreiter1997} with gating mechanism to handle the information flow. LSTM cannot handle longer sequences but it had a main problems. 
 Another proposal came from natural language processing field \cite{vaswani2017attention}. This was a sequence-to-sequence model based entirely on attention mechanism dispensing with recurrence and convolutions. for similar input sequence the model computes contextualized representations using \textit{multi-head self-attention} operating on the output of a position-wise feedforward layers. For each attention head, the input is linearly projected into queries, keys, and values:
\begin{equation}
	\mathbf{Q} = \mathbf{X} \mathbf{W}_Q, \quad
	\mathbf{K} = \mathbf{X} \mathbf{W}_K, \quad
	\mathbf{V} = \mathbf{X} \mathbf{W}_V,	
	\label{transformer}
\end{equation}

where $\mathbf{W}_Q, \mathbf{W}_K, \mathbf{W}_V \in \mathbb{R}^{d \times d_h}$ are learned projection matrices and $d_h = d / h$ for $h$ heads. The scaled dot-product attention for each head is computed as:
\[
\text{Attention}(\mathbf{Q}, \mathbf{K}, \mathbf{V}) = \text{softmax}\left(\frac{\mathbf{Q} \mathbf{K}^\top}{\sqrt{d_h}}\right)\mathbf{V}.
\]
This requires computing a similarity matrix $\mathbf{Q}\mathbf{K}^\top \in \mathbb{R}^{T \times T}$ for every attention head, leading to a time and space complexity of $\mathcal{O}(T^{2d})$ per layer. This quadratic complexity with respect to sequence length $T$ is a key limitation of standard self-attention mechanisms, especially for long sequences.

The outputs of all $h$ heads are concatenated and projected:
\[
\text{MultiHead}(\mathbf{X}) = \text{Concat}(\text{head}_1, \dots, \text{head}_h) \mathbf{W}^O,
\]
with $\mathbf{W}^O \in \mathbb{R}^{d \times d}$. This is followed by a residual connection and layer normalization:
\[
\mathbf{Z}^{(1)} = \text{LayerNorm}(\mathbf{X} + \text{MultiHead}(\mathbf{X})).
\]
Then, a position-wise feedforward network is applied:
\[
\text{FFN}(\mathbf{z}) = \text{ReLU}(\mathbf{z} \mathbf{W}_1 + \mathbf{b}_1)\mathbf{W}_2 + \mathbf{b}_2,
\]
followed again by residual connection and normalization:
\[
\mathbf{Z}^{(2)} = \text{LayerNorm}(\mathbf{Z}^{(1)} + \text{FFN}(\mathbf{Z}^{(1)})).
\]
Despite of quadratic computational complexity, attentions can be computed in parallel on GPUs. such parallel computations were not possible in recurrent based architectures. there were many proposals for reducing the complexity of Transformers for example by making sparse the dense matrices or other mathematical tricks.

Although, attention mechanism calculates correlations of all pairs of words and stores them in state. Therefore, any attempt to reduce the size of state in attention mechanism will result in performance loss. 

\subsection{State Space Models}\label{SSM:exp}

State Space Models (SSMs) describe the evolution of hidden states over time using linear dynamical systems and can be viewed as a continuous generalization of RNNs. Let $\mathbf{h}_t \in \mathbb{R}^n$ be the hidden state and $\mathbf{x}_t \in \mathbb{R}^m$ the input at time $t$. In discrete form, the standard linear SSM is defined as:
\begin{align}
	\label{dynamic:eq}
\mathbf{h}_{t+1} = \mathbf{A} \mathbf{h}_t + \mathbf{B} \mathbf{x}_t, \quad \mathbf{y}_t = \mathbf{C} \mathbf{h}_t + \mathbf{D} \mathbf{x}_t,
\end{align}
where $\mathbf{A}, \mathbf{B}, \mathbf{C}, \mathbf{D}$ are learnable affine Transformation. This recurrence is equivalent to RNNs without nonlinearity mapping function.

from a dynamical systems perspective, the continuous-time analog of this model corresponds to linear time-invariant (LTI) system governed by an ordinary differential equation:
\begin{align}
\frac{d\mathbf{h}(t)}{dt} = \mathbf{A} \mathbf{h}(t) + \mathbf{B} \mathbf{x}(t), \quad \mathbf{y}(t) = \mathbf{C} \mathbf{h}(t) + \mathbf{D} \mathbf{x}(t),
\end{align}

which, when discretized (using methods like Bi-Linearization or Rung-Kutta), yields the earlier recurrence. This shows that discrete SSMs arise naturally from continuous-time physics-based models.

Similar to the pervious models, the discrete dynamics over time illustrates a the linear temporal structure:

\begin{align}
	\mathbf{h}_1 &= \mathbf{A} \mathbf{h}_0 + \mathbf{B} \mathbf{x}_0, 
	\nonumber \\
	\mathbf{h}_2 &= \mathbf{A}^2 \mathbf{h}_0 + \mathbf{A} \mathbf{B} \mathbf{x}_0 + \mathbf{B} \mathbf{x}_1, \nonumber \\ 
	\mathbf{h}_3 &= \mathbf{A}^3 \mathbf{h}_0 + \mathbf{A}^2 \mathbf{B} \mathbf{x}_0 + \mathbf{A} \mathbf{B} \mathbf{x}_1 + \mathbf{B} \mathbf{x}_2.\nonumber \\ 
	\dots \nonumber \\
	\mathbf{h}_T &= \mathbf{A}^T \mathbf{h}_0 + \mathbf{A}^{T-1} \mathbf{B} \mathbf{x}_0 + \mathbf{A}^{2} \mathbf{B} \mathbf{x}_1 +  \dots + \mathbf{B} \mathbf{x}_2. 
	\label{unfolded SSM:eq}
\end{align}

which is a linear convolution of past inputs with the sequence $\{\mathbf{A}^T,\mathbf{A}^{T-1}, \mathbf{A}^{T-2}, \mathbf{A}^{T-3} \dots, \mathbf{A}\}$ as a temporal filter.

This convolution representation of the SSM helps much in lowering computational complexity.

Because of linearity and high parameter sharing, SSMs are highly parallelizable especially when $\mathbf{A}$ A is  diagonal or low-rank This enables efficient long-range sequence modeling, which is mandatory in modern architectures, and avoid memory and gradient bottlenecks of traditional RNNs.

\subsection{Special Structure Matrices}\label{SpecStr:exp}
SSM structure in 3 requires four transformations of $\mathbf{A},\mathbf{B},\mathbf{C},\mathbf{D}$. Specially square matrix $\mathbf{A}$ plays substantial role in remembering information, with learnable parameters. but even with lower complexity compared to In first generation S4 model the structure of $\mathbb{A}$ matrices was HiPPo \cite{Gu2020} significantly improving the performance with respect to  unstructured state design SSM models. But there was another problem; Clearly in \ref{unfolded SSM:eq}, for longer time step sequences, high powers of  $\mathbf{A}$ matrices must be calculated, which is hard even in Hippo. To mitigate the issue, in S5 a diagonal plus low rank structure was proposed. In the last generation S6 \cite{Gu2022}, a totally diagonal structure was used.

\subsection{Mamba} \label{mamba:exp}
Mamba architecture \cite{gu2023mamba} is a functional SSM, with high capacity to model long-term dependencies.
Mamba is an SSM module wrapped in a Gated Multi-Layer Perceptron (G-MLP) resembling a transformer unit. 
G-MLP has two paths: First, which map the input data by Linear projection, 1-D convolution, and an SSM processor with Diagonal State Space (DSS) Matrix a nonlinear activation function. Second path is a skip connection with a nonlinear activation function and multiplied with the first path.
Finally it goes through another MLP network to the output.

Given an input sequence $\mathbf{X} = [\mathbf{x}_0, \dots, \mathbf{x}_{T-1}] \in \mathbb{R}^{d_{\text{in}}\times T}$, Mamba computes a hidden state sequence $\mathbf{H} = [\mathbf{h}_0, \dots, \mathbf{h}_{T-1}] \in \mathbb{R}^{d \times T}$ via the recurrence:
\[
\mathbf{h}_{t+1} = \mathbf{A} \mathbf{h}_t + \mathbf{B} (\mathbf{x}_t \odot \mathbf{g}^{\text{in}}_t)
\]
\[
\mathbf{y}_t = \mathbf{o}_t \odot (\mathbf{C} \mathbf{h}_t)
\]
where: $\mathbf{g}^{\text{in}}_t = \text{Gate}^{\text{in}}(\mathbf{x}_t)$ is the input modulation gate, $\mathbf{o}_t = \text{Gate}^{\text{out}}(\mathbf{x}_t)$ is the output gating vector, $\odot$ denotes element-wise vector multiplication.

These gates are typically implemented using small MLPs or linear projections followed by activations like $\sigma$ (e.g., $\text{SiLU}$). This allows the model to control both how inputs affect the state and how the state affects output.

In practice, Mamba replaces the sequential recurrence with a convolutional formulation derived from state evolution equation:
\[
\mathbf{y} = K * (\mathbf{x} \odot \mathbf{g}^{\text{in}}) \odot \mathbf{g}^{\text{out}}
\]
where $K \in \mathbb{R}^{L \times d}$ is a parameterized kernel derived from discretized state transition matrix $\mathbf{A}$, and $*$ denotes 1D convolution across time. The output gate $\mathbf{g}^{\text{out}}$ modulates the final output.

\textbf{Computational Complexity:} Unlike Transformers, which compute pairwise attention with complexity $\mathcal{O}(T^2 \cdot d)$, Mamba achieves \textit{linear} complexity may be time or space $\mathcal{O}(T \cdot d)$ by:
1) Avoiding pairwise softmax attention,
2) Using efficient convolution or \texttt{scan}-based recurrence implementations,
3)sharing the kernel across tokens.

This enables Mamba to scale to long contexts (e.g., $T > 100{,}000$) with low memory overhead, while still modeling long-range dependencies through the structured kernel $K$ and learned gates.

\section{Quantum overview}
\subsection{Loading classical data} \label{encoding:exp}
Since VQCs operate on quantum states, which lives in Hilbert space; classical data in lowest computational basis (i.e., $0$ and $1$) must load to quantum registers and simulate their information as an element in Hilbert space.
There are multiple kinds of data loading, such as basis encoding, angle encoding with qubit rotational gates and amplitude encoding. Here we use amplitude encoding where classical information is embedded into the amplitudes of a quantum state.
Given a classical matrix $\mathbf{x} \in \mathbb{R}^{2^n\times 2^n}$, the vector is normalized and encoded as a quantum state over $n$ qubits:
\[
\ket{S_{\mathbf{x}}} = \sum_{i=0}^{2^n - 1} x_i \ket{i}, \quad \text{where} \quad \sum_{i=0}^{2^n - 1} |x_i|^2 = 1.
\]
This representation allows one to encode $2^n$ classical features using only $n$ qubits, leading to exponential compression of classical information.

As an example, a sequence $\mathbf{x} \in \mathbb{R}^{4\times 2}$ spread $T = 4$ time steps,with $d = 2$ features; if we flatten it into a vector:
\[
\mathbf{x} = [x_0^{(0)}, x_1^{(0)}, x_0^{(1)}, x_1^{(1)}, x_0^{(2)}, x_1^{(2)}, x_0^{(3)}, x_1^{(3)}] \in \mathbb{R}^8.
\]
We then normalize this vector and encode it as a quantum state over 3 qubits:
\begin{align}
\ket{S_{\mathbf{x}}} = \sum_{i=0}^{7} x_i \ket{i}, \quad \text{with} \quad \sum_{i=0}^7 |x_i|^2 = 1.
\end{align}
However there are other alternatives that consider correlations of feature dimension describing a certain time sequence. those techniques are often used in image data encoding modality.

Each computational basis state $\ket{i}$ (for $i \in \{0, \dots, 7\}$) corresponds to a 3-Qubit tensorized basis to describe exponentially large Hilbert space. 

\begin{figure}
	\includegraphics[scale=0.20]{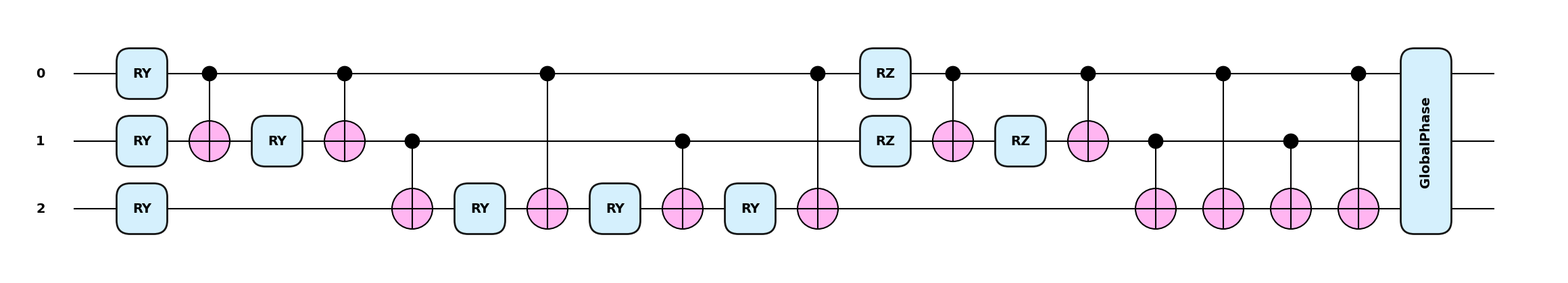}
	\caption{Amplitude encoding block. There are 3 main part for that circuit first, amplitude of classical data which is going to happen by $R_y$ and entangle gates, second loading phase of classical data into quantum state in such case that classical data is complex. Finally adding overall global phase to the whole quantum state}
	\label{ampEncode}
\end{figure}

Amplitude encoding Fig.~\ref{ampEncode},  reduces an exponentially large classical vector to a linear number of qubits. Specifically, to store a classical vector of length $2^n$, a quantum system only needs $n$ qubits. This leads to exponential savings in memory and opens the door to quantum algorithms that can process large datasets in logarithmic time with respect to input size.

\subsection{Ansatz considerations} \label{ansatz:exp}

Since we picked amplitude encoding for data loading, Next, we should choose compatible ansatz which gives proper expressivity and entangling capability. in \cite{sim2019expressibility} there are 17 series of circuits designed using entanglement gates or entangling instructions via nearest neighbour, circuit block, or all-to-all in expressivity and entangling capability domain. We provide our circuits on Pennylane framework \cite{bergholm2022}. Also \cite{schuld2020circuit} proposed a maximally entangled layer structure used for classification problem which is used here as our Quantum circuit part.

Assume $U_{\text{Q}}(\boldsymbol{\theta})$ is a parameterized complex unitary entangling quantum gate  that act on a quantum state :

\begin{equation}
	\big|\phi(x;\boldsymbol{\theta})\big\rangle
	\;=\; U_{\text{Q}}(\boldsymbol{\theta}) \, \big|\psi(x)\big\rangle.
\end{equation}

However, each composite gate $U_{\text{Q}}(\boldsymbol{\theta})$ internally uses two types of gates: parameterized rotation and quantum entangling. 
\begin{equation}
	U_{\text{Q}}(\boldsymbol{\theta}) \;=\; \prod_{\ell=1}^{L_Q} \Big( U_{\text{ent}}^{(\ell)} \, U_{\text{rot}}^{(\ell)}(\boldsymbol{\theta}^{(\ell)}) \Big).
\end{equation}

\begin{figure}[H]
	\includegraphics[scale=0.4]{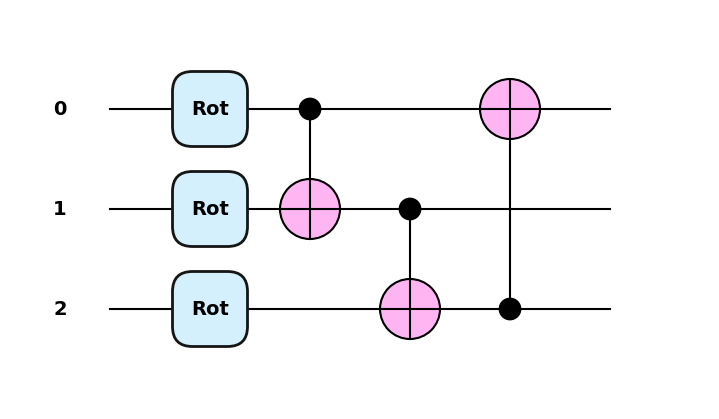}
	\centering\caption{A maximally entangling composite gate applied to a 3-qubit space. Each $Rot$ gate applied to the qubits consists of three serial gates $\alpha,\beta,\gamma$. }
	\label{strongEnt}
\end{figure}

Examples of highly entangling structures Fig.~\ref{strongEnt}, include repeated layers of single-qubit rotations followed by an all-to-all pattern of controlled-phase/CZ gates, or a GHZ-style entangling layer of a Hadamard on one qubit followed by cascading CNOTs. In general, we do not require a specific circuit; we only need the unitary operator to generate non-trivial multi-qubit correlations.

\subsubsection{Expressivity metric} \label{expressibility:exp}
Expressivity metric indicates how much a generated quantum state by a circuit is well estimated from haar distribution. Consider a finite set of generated quantum states $\set{\ket{\psi_i}}_{i=1}^n$, by a VQC circuit. if 2 states arbitrary state $\ket{\psi_\phi}$ and $\ket{\psi_\theta}$ are drawn randomly and independently the fidelity is defined as similarity between those states:
\begin{align}
	F := |\langle \psi_\theta | \psi_\phi \rangle|^2
\end{align} 
 so each time this fidelity computes a correlation between 2 state vector. Fidelity is itself a random variable with distribution $\mathbf{P(F)}$.\\
Fidelity for a Haar-distibuted random state.

\begin{equation}
	P_{\text{Haar}}(F) = (N - 1)(1 - F)^{N - 2}
\end{equation}

Haar distribution is unitary-invariant, This means if any states is drawn such $\ket{\psi}\sim \mathbf{P_{Haar}}$, then a unitary gate $\mathbf{U}$ won't change the distribution.

The $t$-th order \textbf{frame potential} is defined as the t-th moment of fidelity distribution by:
	
\begin{align}
	F^{(t)} &= \int_{\Theta} \int_{\Phi} |\langle \psi_\theta | \psi_\phi \rangle|^{2t} \, d\theta \, d\phi \\
	&= \mathbb{E}_\theta \mathbb{E}_\phi \left[ |\langle \psi_\theta | \psi_\phi \rangle|^{2t} \right] \nonumber \\
	&= \mathbb{E}[F^t]
\end{align}

The expressivity of a PQC is the KL distance between fidelity distribution from the PQC and Haar distribution:
	
\begin{equation}
	\text{Expr} = D_{\text{KL}}\left( \hat{P}_{\text{PQC}}(F; \theta) \,\|\, P_{\text{Haar}}(F) \right)
\end{equation}

The upper bound on expressibility for a non-expressive circuit (e.g., fixed output state) is:
\begin{equation}
	\text{Expr}_{\max} = (N - 1) \ln(n_{\text{bin}})
\end{equation}

In general, expressivity of a VQC would tell about the efficient power of exploring randomness over Hilbert space.
Low expressivity means VQC's insufficient coverage of Hilbert-space with poor generalization and underfitting.\\
High expressivity describes that circuit covers a large Hilbert space as solution definition but it overfit the QML model if the input data be less enough.
Then expressivity metric is a critical option which must determines for reaching best accuracy in model.  

\section{our work} \label{ourwork:exp}

\begin{figure}[H]
	\includegraphics[scale=0.3]{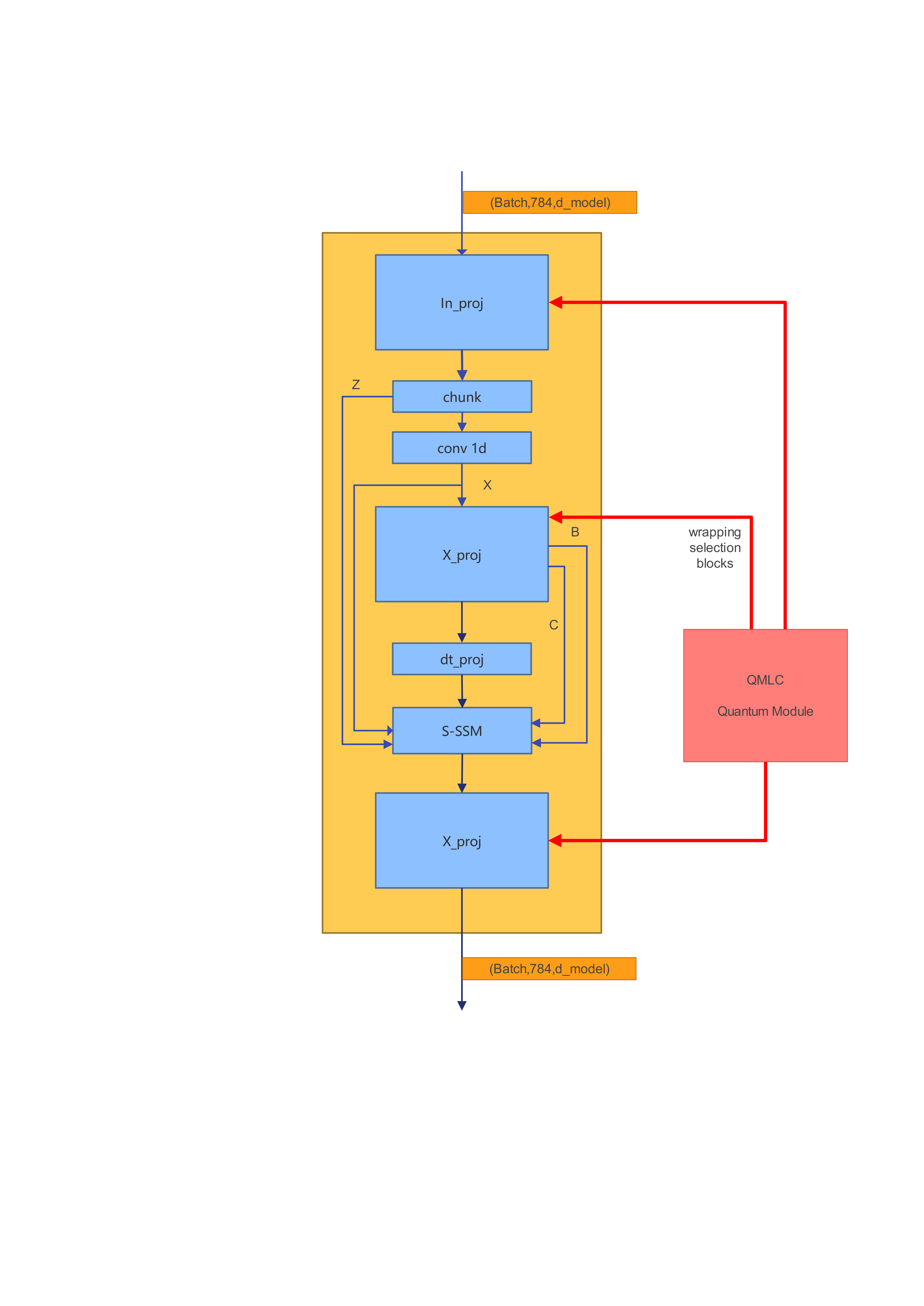}
	\caption{The proposed hybrid quantum-classical architecture. Three projection mechanisms (\texttt{in\_proj}, \texttt{x\_proj}, and \texttt{out\_proj}) are wrapped by the quantum transformation modules presented in Sections~\ref{encoding:exp} and~\ref{ansatz:exp}.}
	\label{QmambaArch}
\end{figure}

In this work, we propose a hybrid quantum-classical architecture Fig.~\ref{QmambaArch}, that integrates a Variational Quantum Circuit with amplitude encoding (as described in Section~\ref{encoding:exp}) and a maximally entangling layer (introduced in Section~\ref{ansatz:exp}). The quantum module acts as a \textit{Quantum Transformation Block}, effectively replacing three key classical projection components in the Mamba architecture. These components are described as follows:

\begin{enumerate}
	\item \textbf{in\_proj:} This projection takes an input tensor of shape $(\text{batch}, \text{length}, d_{\text{model}})$ and outputs $(\text{batch}, \text{length}, 2 \times d_{\text{inner}})$. Its purpose is to project the input data into the latent representation space of Mamba, enabling richer feature encoding.
	
	\item \textbf{x\_proj:} This projection operates on the output of the preceding convolutional block, receiving $(\text{batch}, \text{length}, d_{\text{inner}})$ and returning $(\text{batch}, \text{length}, 2 \times d_{\text{state}} + d_{t\_\text{rank}})$. It prepares the input and output representations for the selective state-space scan, while also encoding time-dependent rank information ($d_{t\_\text{rank}}$).
	
	\item \textbf{out\_proj:} This block processes $(\text{batch}, \text{length}, d_{\text{inner}})$ representations obtained from the selective state-space module and maps them back to the original dimensionality $(\text{batch}, \text{length}, d_{\text{model}})$, restoring the output format compatible with the overall model architecture.
\end{enumerate}

dt\_proj transformation, which maps $(batch,length,dt\_rank)$ to $(batch,length, d\_inner)$, is retained as a classical component. We observed during numerical simulation that this module is highly sensitive, as it projects the $delta\_t$ information used for discretizing both input and memory-state dynamics, represented by $A$ matrix in the selective mechanism. Therefore, We leaved this selection method to act as a classical projection. 

This hybrid configuration was compared with a fully classical model employing traditional selection mechanisms, using reshaped MNIST dataset as a benchmark. The comparison highlights the effectiveness and expressivity of the quantum-enhanced transformation under identical training conditions, particularly in tasks involving complex classification boundaries. Fig.~\ref{},
Data shape in MNIST is $(batch,1,32,32)$ as standard gray scale image dataset structure. after reshaping its shape is $(batch,784,1)$.
For more effective sequence processing, the dimension of the model embedding $d\_model$ must be sufficiently large. Then, we increase $d\_model$ from 1 to 128 for the Hybrid Quantum Selection mechanism. In the purely classical configuration, $d\_model$ is raised to 16 to mitigate overfitting during small-scale training and to ensure a fair comparison in terms of parameter count.

\subsection{optimization process} \label{optim:exp}
In both classical and VQC implementations of Mamba structure, optimizing model parameters to reach a satisfactory performance is the main objective.
In the classical domain we encounter the vanishing or exploding gradient phenomena which hinders good training of model parameters. 
In the quantum domain, we face a similar difficulty of Barren-plateaus \cite{mcclean2018}.
We separated parameters of each Quantum Selection mechanism and set different strategy to optimize hyper-parameters individually.
All classical sections such as mamba block and other modules \cite{paszke2019PyTorch}. This framework enables us to group All those parameters by their characteristic. In Hybrid Quantum Classical selection mechanism we set learning rates to 3e-4, 1e-4, 3e-4 to in\_proj, x\_proj, out\_proj respectively. All other classical parameters considered as 1e-3. Also weight\_decay considered as 0.01 for Classical parameters and zero for all Selection mechanisms, due to the sensitivity of Quantum parameters.

\section{results}
\begin{figure}[H]
	\includegraphics[scale=0.3]{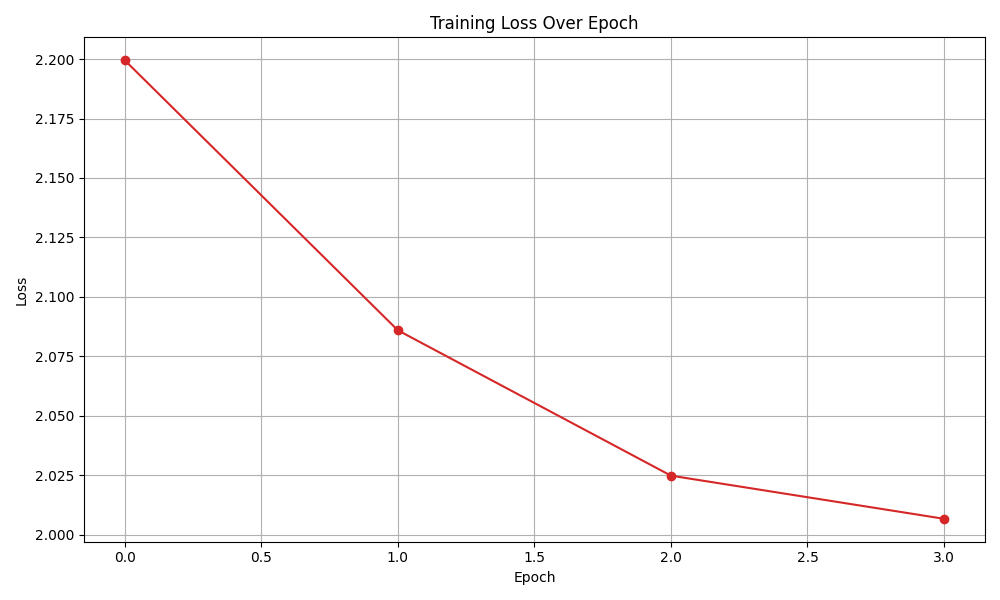}
	\caption{Average Loss over 4 epochs for HQC selection mamba}
	\label{AvgLHQC}
\end{figure}

\begin{figure}[H]
	\includegraphics[scale=0.3]{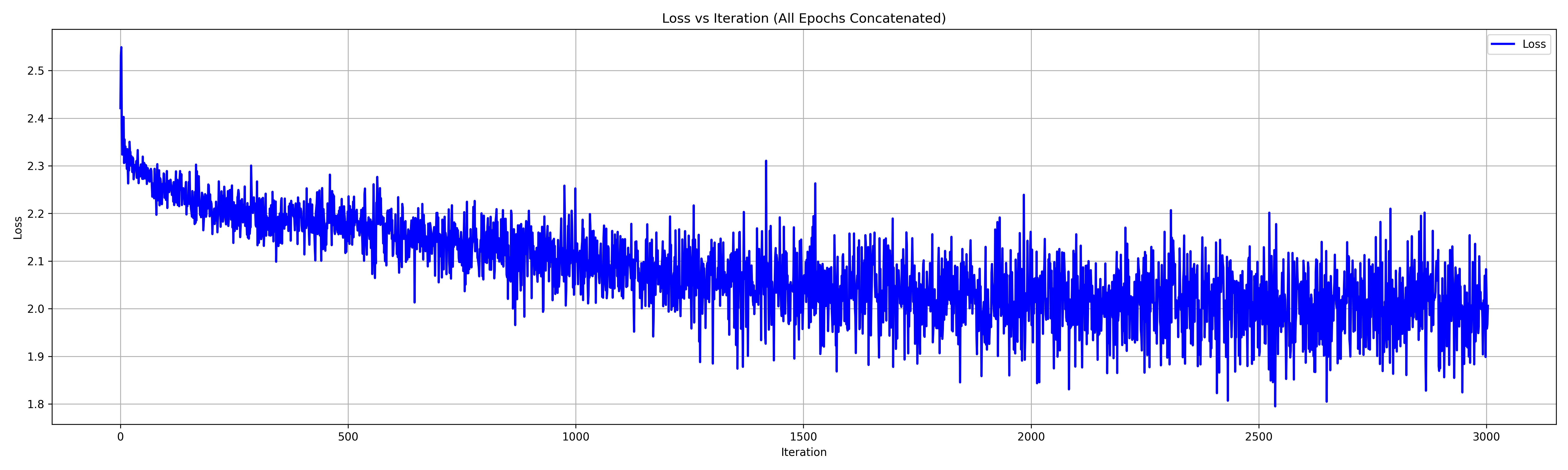}
	\caption{Loss versus iterations during 4 epochs.}
	\label{LossItrHQC}
\end{figure}

\begin{figure}[H]
	\includegraphics[scale=0.3]{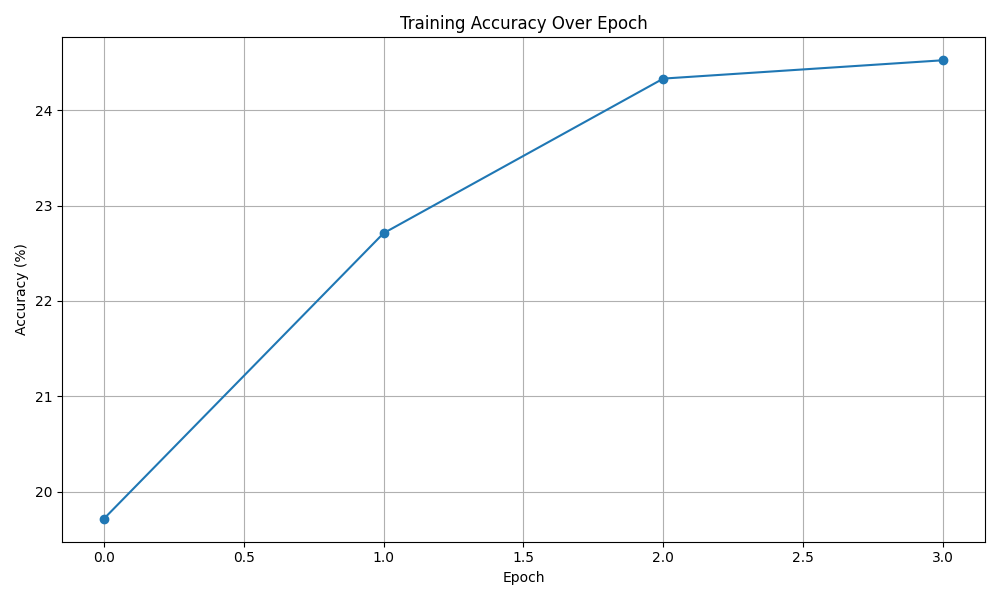}
	\caption{Average Accuracy per all iterations during 4 epochs.}
	\label{acc_itr_HQC}
\end{figure}

\begin{figure}[H]
	\includegraphics[scale=0.3]{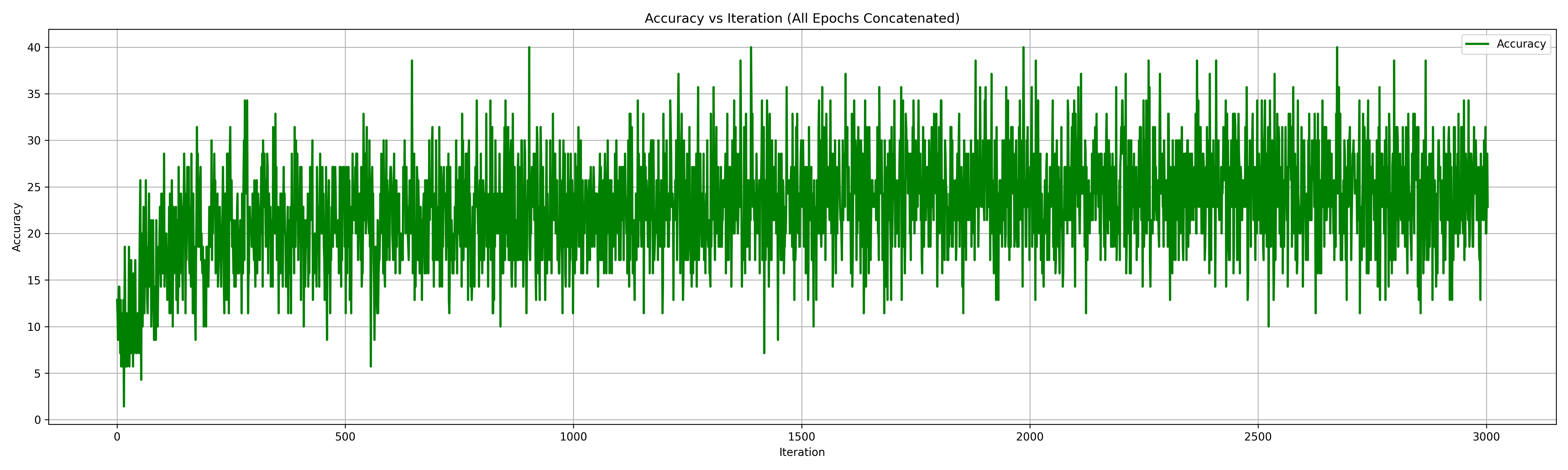}
	\caption{Accuracy versus per all iterations during 4 epochs.}
	\label{acc_itr_HQC_noavg}
\end{figure}

\begin{figure}[H]
	\includegraphics[scale=0.3]{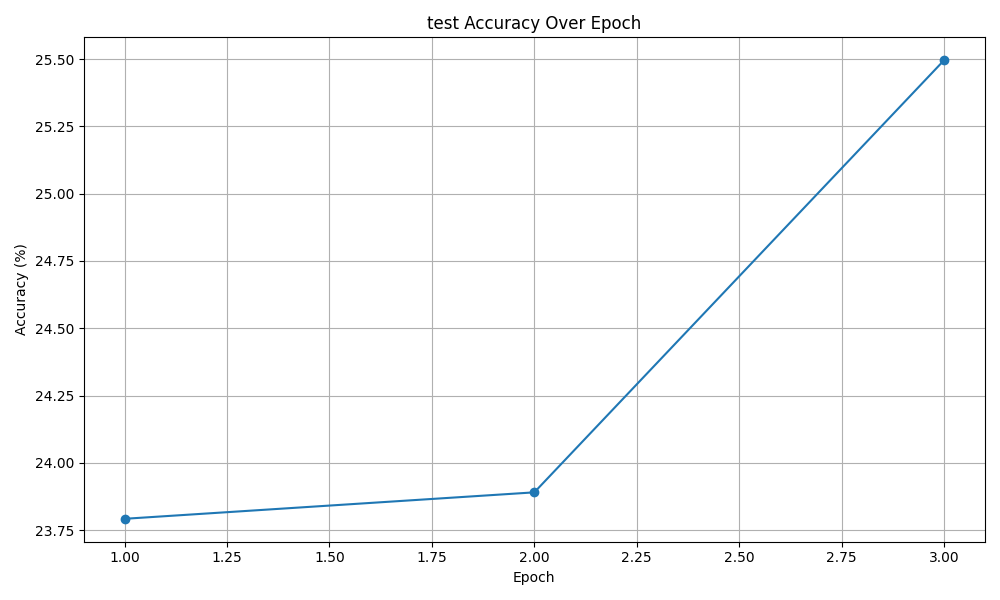}
	\caption{Test accuracy iterations during 4 epochs.}
	\label{acc_test_HQC}
\end{figure}

\begin{figure}[H]
	\includegraphics[scale=0.3]{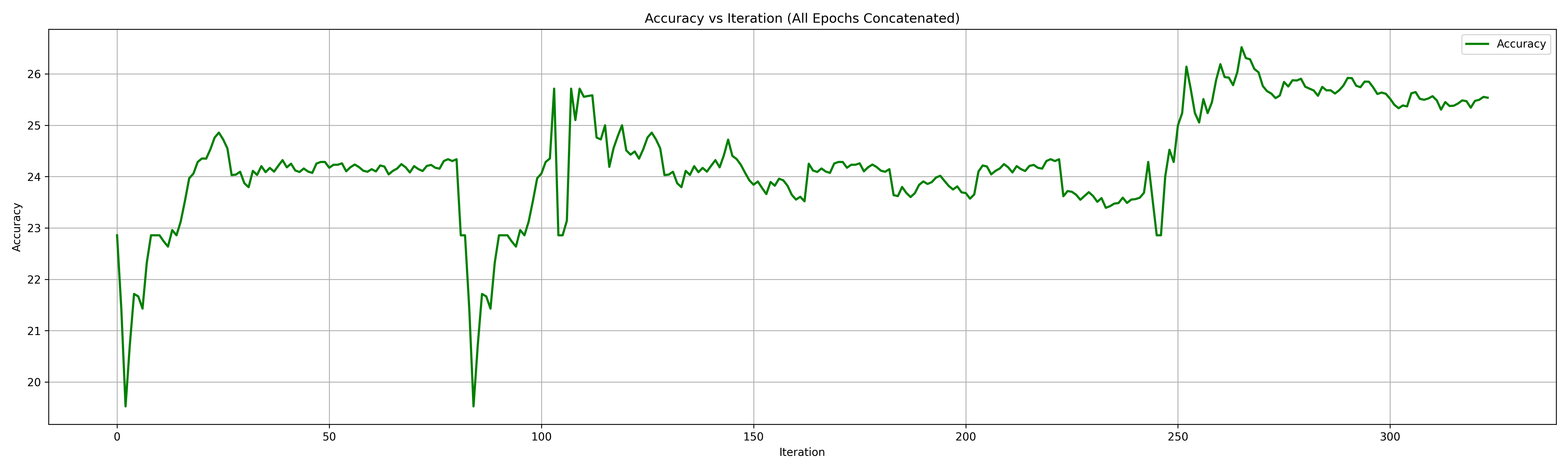}
	\caption{Test accuracy versus iterations during 4 epochs.}
	\label{acc_test_HQC_noavg}
\end{figure}

Our flatten MNIST dataset designed to illustrates the effectiveness of expressivity VQC promised against classical projectors.
Since we had limitation of computational resources this scheme where suggested to show that advantage in initial epochs. Each epoch with full size of MNIST dataset takes around 1.20 hours with proposed structure. Therefore, we processed up in to 4 epochs both classical and HQC proposal structure and here is the results.
d\_model is 128 with 2 layer HQC selection mamba and expanding rate 4 and d\_state 16.
with that config Fig.~\ref{AvgLHQC}, shows the Average Loss per 4 epochs. Total loss during all epochs indicated in Fig.~\ref{LossItrHQC}. Average accuracy in HQC case presented by Fig.~\ref{acc_itr_HQC}, Fig.~\ref{acc_itr_HQC_noavg}, in average and total respectively. Similarly test process figure indicated in Fig.~\ref{acc_test_HQC}, and Fig.~\ref{acc_test_HQC_noavg}

\begin{figure}[H]
	\includegraphics[scale=0.3]{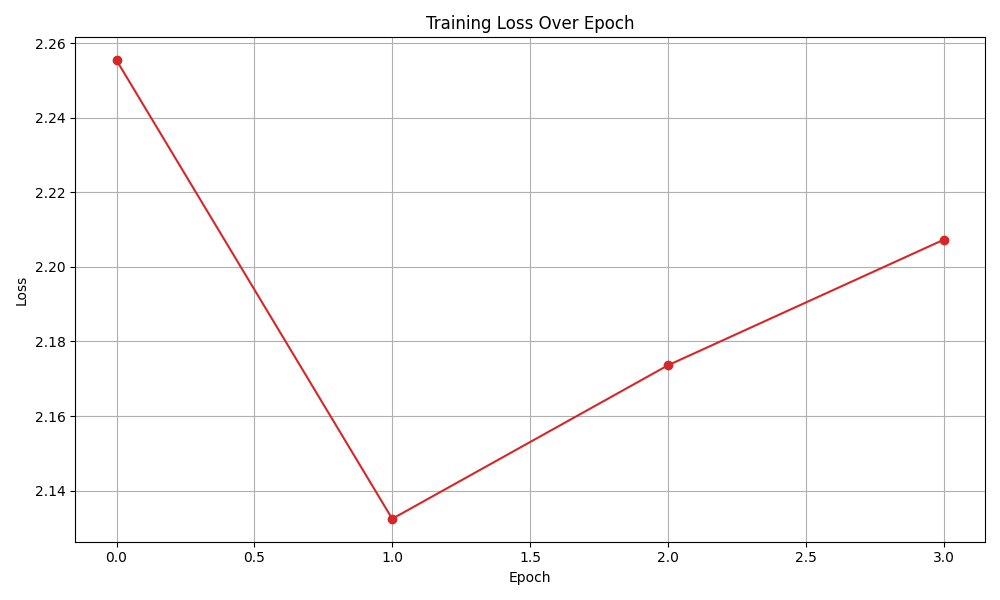}
	\caption{Average loss over 4 epochs for HQC selection mamba}
	\label{AVGLC}
\end{figure}

\begin{figure}[H]
	\includegraphics[scale=0.3]{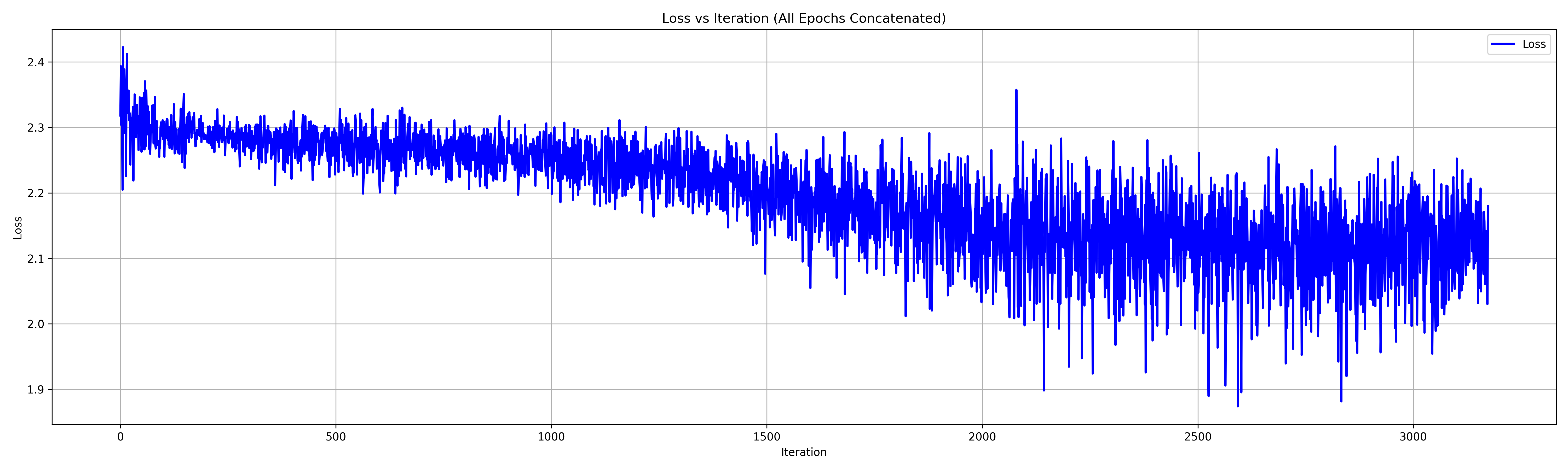}
	\caption{Overall Loss per all iterations during 4 epochs.}
	\label{LC}
\end{figure}

\begin{figure}[H]
	\includegraphics[scale=0.3]{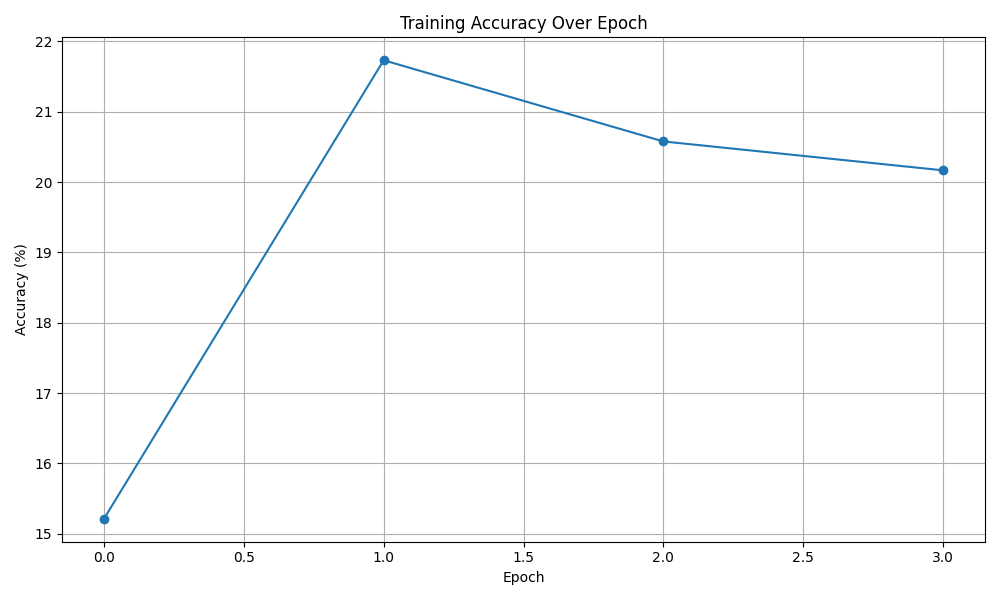}
	\caption{Loss versus iterations during 4 epochs.}
	\label{acc_itr_c}
\end{figure}

\begin{figure}[H]
	\includegraphics[scale=0.3]{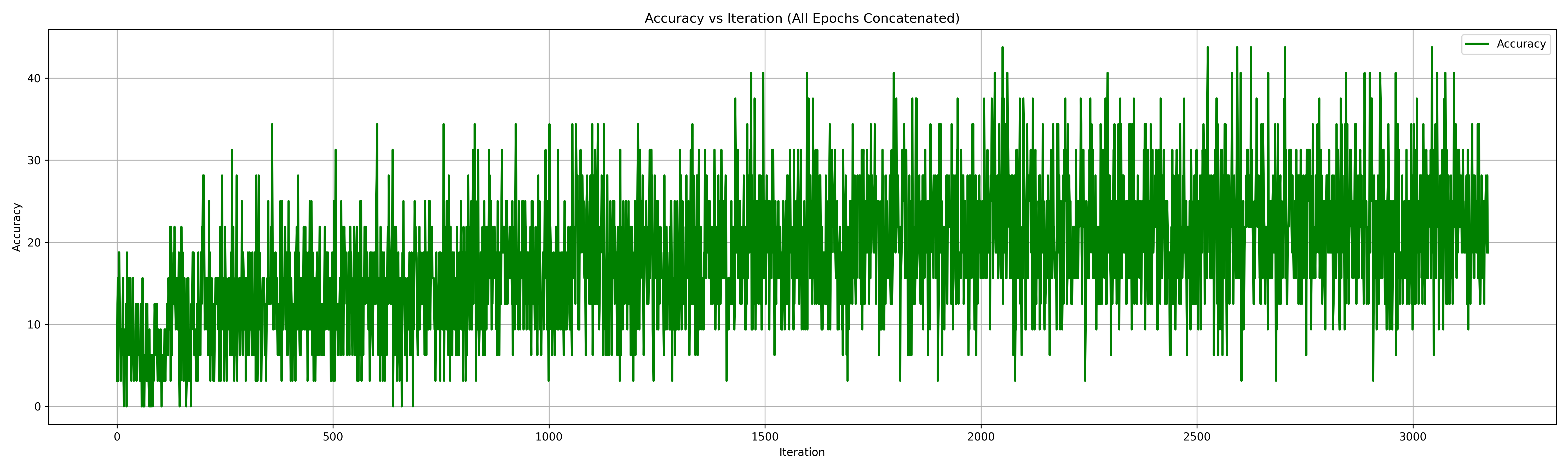}
	\caption{Overall accuracy per all iterations during 4 epochs.}
	\label{acc_itr_OC}
\end{figure}

\begin{figure}[H]
	\includegraphics[scale=0.3]{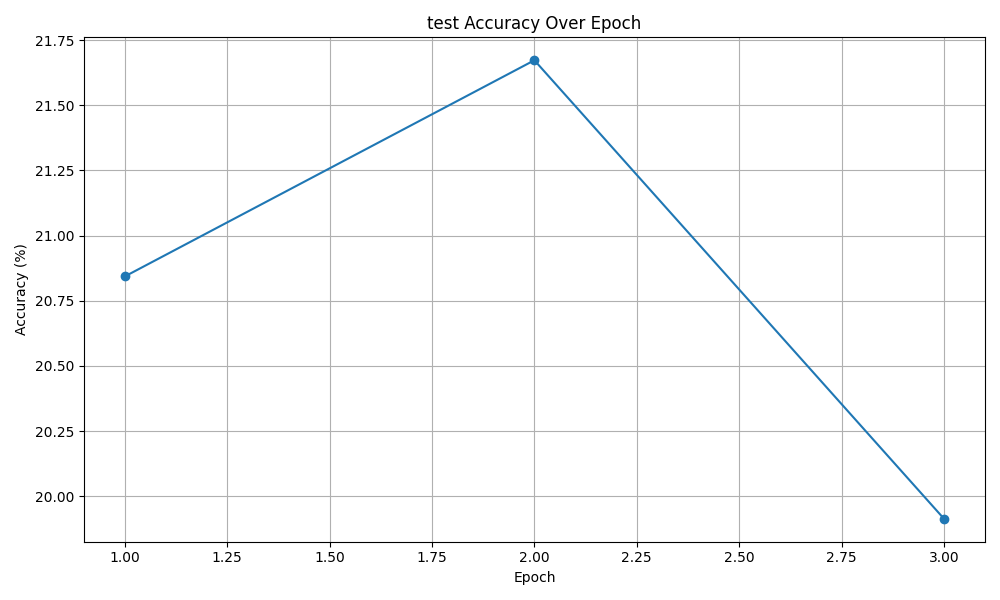}
	\caption{Overall test per all iterations during 4 epochs.}
	\label{acc_test_c}
\end{figure}

\begin{figure}[H]
	\includegraphics[scale=0.3]{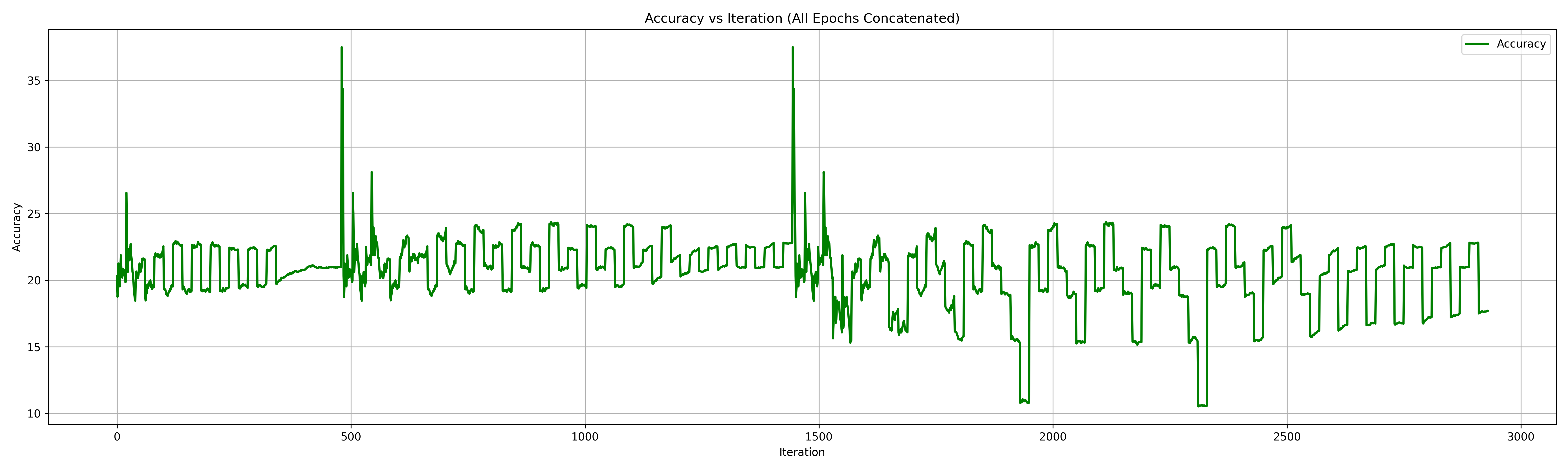}
	\caption{Overall test per all iterations during 4 epochs.}
	\label{acc_test_OC}
\end{figure}

For Classical Case, the results of loss indicates in Fig.~\ref{AVGLC}, Fig.~\ref{LC} and accuracy of that scheme showed in Fig.~\ref{acc_itr_c},Fig.~\ref{acc_itr_OC}. Also the test figures presented in Fig.~\ref{acc_test_c}, Fig.~\ref{acc_test_OC}.

Hence we reduced $d\_model=16$ to prevent overfiting. However, under this configuration, loss was increased and accuracy reduced after first epoch Besides, 
increasing the model size leads to severe overfitting given the limited training data.

If we reduce the size of parameters, the model couldn't efficiently extract all those features from a reshaped 784 elements and it failed to learn.
 
Therefore, a trade-off exists between overfitting and underfitting, representing a narrow threshold at which the model can maximally capture the underlying semantic structure of the data.

In the classical Mamba model \ref{acc_test_classical}, the maximum test accuracy was achieved in the second epoch, reaching 21.7\%, after which it dropped below 20\% in subsequent epochs. In contrast, the averaged-iteration procedure \ref{acc_itr_classical} achieved the same peak accuracy in the second epoch but maintained it more consistently in later epochs compared to the test case.
This observation suggests that the classical model exhibits limited robustness and poor generalization capability. 

In comparison, our proposed Hybrid Quantum–Classical (HQC) model during same 4 epochs training steps \ref{acc_itr_HQC} achieved its highest accuracy of 24.7\% in the final epoch after four training epochs. Unlike the classical model, the HQC model maintained a more stable and significant accuracy trend across both training and test datasets \ref{acc_test_HQC}, demonstrating improved generalization performance.

\section*{Funding}
This research received no external funding.

\bibliography{sn-bibliography}


\begin{thebibliography}{46}
\ifx \bisbn   \undefined \def \bisbn  #1{ISBN #1}\fi
\ifx \binits  \undefined \def \binits#1{#1}\fi
\ifx \bauthor  \undefined \def \bauthor#1{#1}\fi
\ifx \batitle  \undefined \def \batitle#1{#1}\fi
\ifx \bjtitle  \undefined \def \bjtitle#1{#1}\fi
\ifx \bvolume  \undefined \def \bvolume#1{\textbf{#1}}\fi
\ifx \byear  \undefined \def \byear#1{#1}\fi
\ifx \bissue  \undefined \def \bissue#1{#1}\fi
\ifx \bfpage  \undefined \def \bfpage#1{#1}\fi
\ifx \blpage  \undefined \def \blpage #1{#1}\fi
\ifx \burl  \undefined \def \burl#1{\textsf{#1}}\fi
\ifx \doiurl  \undefined \def \doiurl#1{\url{https://doi.org/#1}}\fi
\ifx \betal  \undefined \def \betal{\textit{et al.}}\fi
\ifx \binstitute  \undefined \def \binstitute#1{#1}\fi
\ifx \binstitutionaled  \undefined \def \binstitutionaled#1{#1}\fi
\ifx \bctitle  \undefined \def \bctitle#1{#1}\fi
\ifx \beditor  \undefined \def \beditor#1{#1}\fi
\ifx \bpublisher  \undefined \def \bpublisher#1{#1}\fi
\ifx \bbtitle  \undefined \def \bbtitle#1{#1}\fi
\ifx \bedition  \undefined \def \bedition#1{#1}\fi
\ifx \bseriesno  \undefined \def \bseriesno#1{#1}\fi
\ifx \blocation  \undefined \def \blocation#1{#1}\fi
\ifx \bsertitle  \undefined \def \bsertitle#1{#1}\fi
\ifx \bsnm \undefined \def \bsnm#1{#1}\fi
\ifx \bsuffix \undefined \def \bsuffix#1{#1}\fi
\ifx \bparticle \undefined \def \bparticle#1{#1}\fi
\ifx \barticle \undefined \def \barticle#1{#1}\fi
\bibcommenthead
\ifx \bconfdate \undefined \def \bconfdate #1{#1}\fi
\ifx \botherref \undefined \def \botherref #1{#1}\fi
\ifx \url \undefined \def \url#1{\textsf{#1}}\fi
\ifx \bchapter \undefined \def \bchapter#1{#1}\fi
\ifx \bbook \undefined \def \bbook#1{#1}\fi
\ifx \bcomment \undefined \def \bcomment#1{#1}\fi
\ifx \oauthor \undefined \def \oauthor#1{#1}\fi
\ifx \citeauthoryear \undefined \def \citeauthoryear#1{#1}\fi
\ifx \endbibitem  \undefined \def \endbibitem {}\fi
\ifx \bconflocation  \undefined \def \bconflocation#1{#1}\fi
\ifx \arxivurl  \undefined \def \arxivurl#1{\textsf{#1}}\fi
\csname PreBibitemsHook\endcsname

\bibitem[\protect\citeauthoryear{Preskill}{2018}]{preskill2018}
\begin{barticle}
\bauthor{\bsnm{Preskill}, \binits{J.}}:
\batitle{Quantum {{Computing}} in the {{NISQ}} era and beyond}.
\bjtitle{Quantum}
\bvolume{2},
\bfpage{79}
(\byear{2018})
\doiurl{10.22331/q-2018-08-06-79}
\end{barticle}
\endbibitem

\bibitem[\protect\citeauthoryear{Schuld and Killoran}{2019}]{schuld2019quantum}
\begin{barticle}
\bauthor{\bsnm{Schuld}, \binits{M.}},
\bauthor{\bsnm{Killoran}, \binits{N.}}:
\batitle{Quantum machine learning in feature hilbert spaces}.
\bjtitle{Physical review letters}
\bvolume{122}(\bissue{4}),
\bfpage{040504}
(\byear{2019})
\end{barticle}
\endbibitem

\bibitem[\protect\citeauthoryear{Nielsen and Chuang}{2010}]{nielsen2010quantum}
\begin{bbook}
\bauthor{\bsnm{Nielsen}, \binits{M.A.}},
\bauthor{\bsnm{Chuang}, \binits{I.L.}}:
\bbtitle{Quantum Computation and Quantum Information},
p. \bfpage{305}.
\bpublisher{Cambridge University Press},
\blocation{Cambridge, UK}
(\byear{2010})
\end{bbook}
\endbibitem

\bibitem[\protect\citeauthoryear{Beer et~al.}{2020}]{beer2020training}
\begin{barticle}
\bauthor{\bsnm{Beer}, \binits{K.}},
\bauthor{\bsnm{Bondarenko}, \binits{D.}},
\bauthor{\bsnm{Farrelly}, \binits{T.}},
\bauthor{\bsnm{Osborne}, \binits{T.J.}},
\bauthor{\bsnm{Salzmann}, \binits{R.}},
\bauthor{\bsnm{Scheiermann}, \binits{D.}},
\bauthor{\bsnm{Wolf}, \binits{R.}}:
\batitle{Training deep quantum neural networks}.
\bjtitle{Nature communications}
\bvolume{11}(\bissue{1}),
\bfpage{808}
(\byear{2020})
\end{barticle}
\endbibitem

\bibitem[\protect\citeauthoryear{Farhi
  et~al.}{2014}]{farhiQuantumApproximateOptimization2014a}
\begin{botherref}
\oauthor{\bsnm{Farhi}, \binits{E.}},
\oauthor{\bsnm{Goldstone}, \binits{J.}},
\oauthor{\bsnm{Gutmann}, \binits{S.}}:
A {{Quantum Approximate Optimization Algorithm}}.
arXiv
(2014).
\doiurl{10.48550/arXiv.1411.4028}
\end{botherref}
\endbibitem

\bibitem[\protect\citeauthoryear{Cao et~al.}{2019}]{cao2019}
\begin{barticle}
\bauthor{\bsnm{Cao}, \binits{Y.}},
\bauthor{\bsnm{Romero}, \binits{J.}},
\bauthor{\bsnm{Olson}, \binits{J.P.}},
\bauthor{\bsnm{Degroote}, \binits{M.}},
\bauthor{\bsnm{Johnson}, \binits{P.D.}},
\bauthor{\bsnm{Kieferov{\'a}}, \binits{M.}},
\bauthor{\bsnm{Kivlichan}, \binits{I.D.}},
\bauthor{\bsnm{Menke}, \binits{T.}},
\bauthor{\bsnm{Peropadre}, \binits{B.}},
\bauthor{\bsnm{Sawaya}, \binits{N.P.D.}},
\bauthor{\bsnm{Sim}, \binits{S.}},
\bauthor{\bsnm{Veis}, \binits{L.}},
\bauthor{\bsnm{{Aspuru-Guzik}}, \binits{A.}}:
\batitle{Quantum {{Chemistry}} in the {{Age}} of {{Quantum Computing}}}.
\bjtitle{Chemical Reviews}
\bvolume{119}(\bissue{19}),
\bfpage{10856}--\blpage{10915}
(\byear{2019})
\doiurl{10.1021/acs.chemrev.8b00803}
{\href{https://arxiv.org/abs/1812.09976}{{arXiv:1812.09976}}}
{[quant-ph]}
\end{barticle}
\endbibitem

\bibitem[\protect\citeauthoryear{Harrow et~al.}{2009}]{HHL2009}
\begin{barticle}
\bauthor{\bsnm{Harrow}, \binits{A.W.}},
\bauthor{\bsnm{Hassidim}, \binits{A.}},
\bauthor{\bsnm{Lloyd}, \binits{S.}}:
\batitle{Quantum algorithm for linear systems of equations}.
\bjtitle{Phys. Rev. Lett.}
\bvolume{103},
\bfpage{150502}
(\byear{2009})
\doiurl{10.1103/PhysRevLett.103.150502}
\end{barticle}
\endbibitem

\bibitem[\protect\citeauthoryear{{Bravo-Prieto} et~al.}{2023}]{bravo2023}
\begin{barticle}
\bauthor{\bsnm{{Bravo-Prieto}}, \binits{C.}},
\bauthor{\bsnm{LaRose}, \binits{R.}},
\bauthor{\bsnm{Cerezo}, \binits{M.}},
\bauthor{\bsnm{Subasi}, \binits{Y.}},
\bauthor{\bsnm{Cincio}, \binits{L.}},
\bauthor{\bsnm{Coles}, \binits{P.J.}}:
\batitle{Variational {{Quantum Linear Solver}}}.
\bjtitle{Quantum}
\bvolume{7},
\bfpage{1188}
(\byear{2023})
\doiurl{10.22331/q-2023-11-22-1188}
{\href{https://arxiv.org/abs/1909.05820}{{arXiv:1909.05820}}}
{[quant-ph]}
\end{barticle}
\endbibitem

\bibitem[\protect\citeauthoryear{Childs et~al.}{2017}]{childs2017}
\begin{barticle}
\bauthor{\bsnm{Childs}, \binits{A.M.}},
\bauthor{\bsnm{Kothari}, \binits{R.}},
\bauthor{\bsnm{Somma}, \binits{R.D.}}:
\batitle{Quantum algorithm for systems of linear equations with exponentially
  improved dependence on precision}.
\bjtitle{SIAM Journal on Computing}
\bvolume{46}(\bissue{6}),
\bfpage{1920}--\blpage{1950}
(\byear{2017})
\end{barticle}
\endbibitem

\bibitem[\protect\citeauthoryear{Gily{\'e}n et~al.}{2019}]{gilyen2019}
\begin{bchapter}
\bauthor{\bsnm{Gily{\'e}n}, \binits{A.}},
\bauthor{\bsnm{Su}, \binits{Y.}},
\bauthor{\bsnm{Low}, \binits{G.H.}},
\bauthor{\bsnm{Wiebe}, \binits{N.}}:
\bctitle{Quantum singular value transformation and beyond: exponential
  improvements for quantum matrix arithmetics}.
In: \bbtitle{Proceedings of the 51st Annual ACM SIGACT Symposium on Theory of
  Computing},
pp. \bfpage{193}--\blpage{204}
(\byear{2019})
\end{bchapter}
\endbibitem

\bibitem[\protect\citeauthoryear{Arunachalam and de~Wolf}{2017}]{arunachal2017}
\begin{botherref}
\oauthor{\bsnm{Arunachalam}, \binits{S.}},
\oauthor{\bsnm{Wolf}, \binits{R.}}:
A {{Survey}} of {{Quantum Learning Theory}}.
arXiv
(2017).
\doiurl{10.48550/arXiv.1701.06806}
\end{botherref}
\endbibitem

\bibitem[\protect\citeauthoryear{Biamonte et~al.}{2018}]{biamonte2018}
\begin{botherref}
\oauthor{\bsnm{Biamonte}, \binits{J.}},
\oauthor{\bsnm{Wittek}, \binits{P.}},
\oauthor{\bsnm{Pancotti}, \binits{N.}},
\oauthor{\bsnm{Rebentrost}, \binits{P.}},
\oauthor{\bsnm{Wiebe}, \binits{N.}},
\oauthor{\bsnm{Lloyd}, \binits{S.}}:
Quantum {{Machine Learning}}
(2018).
\doiurl{10.1038/nature23474}
\end{botherref}
\endbibitem

\bibitem[\protect\citeauthoryear{Schuld and Killoran}{2018}]{schuld2018c}
\begin{barticle}
\bauthor{\bsnm{Schuld}, \binits{M.}},
\bauthor{\bsnm{Killoran}, \binits{N.}}:
\batitle{Quantum machine learning in feature {{Hilbert}} spaces}.
\bjtitle{Phys. Rev. Lett. 122, 040504 (2019)}
\bvolume{122}(\bissue{4}),
\bfpage{040504}
(\byear{2018})
\doiurl{10.1103/physrevlett.122.040504}
\end{barticle}
\endbibitem

\bibitem[\protect\citeauthoryear{Havlicek et~al.}{2019}]{havlicek2019}
\begin{barticle}
\bauthor{\bsnm{Havlicek}, \binits{V.}},
\bauthor{\bsnm{C{\'o}rcoles}, \binits{A.D.}},
\bauthor{\bsnm{Temme}, \binits{K.}},
\bauthor{\bsnm{Harrow}, \binits{A.W.}},
\bauthor{\bsnm{Kandala}, \binits{A.}},
\bauthor{\bsnm{Chow}, \binits{J.M.}},
\bauthor{\bsnm{Gambetta}, \binits{J.M.}}:
\batitle{Supervised learning with quantum enhanced feature spaces}.
\bjtitle{Nature}
\bvolume{567}(\bissue{7747}),
\bfpage{209}--\blpage{212}
(\byear{2019})
\doiurl{10.1038/s41586-019-0980-2}
{\href{https://arxiv.org/abs/1804.11326}{{arXiv:1804.11326}}}
{[quant-ph]}
\end{barticle}
\endbibitem

\bibitem[\protect\citeauthoryear{Abbas et~al.}{2021}]{abbas2021}
\begin{barticle}
\bauthor{\bsnm{Abbas}, \binits{A.}},
\bauthor{\bsnm{Sutter}, \binits{D.}},
\bauthor{\bsnm{Zoufal}, \binits{C.}},
\bauthor{\bsnm{Lucchi}, \binits{A.}},
\bauthor{\bsnm{Figalli}, \binits{A.}},
\bauthor{\bsnm{Woerner}, \binits{S.}}:
\batitle{The power of quantum neural networks}.
\bjtitle{Nature Computational Science}
\bvolume{1}(\bissue{6}),
\bfpage{403}--\blpage{409}
(\byear{2021})
\doiurl{10.1038/s43588-021-00084-1}
\end{barticle}
\endbibitem

\bibitem[\protect\citeauthoryear{Huang et~al.}{2022}]{huang2022}
\begin{barticle}
\bauthor{\bsnm{Huang}, \binits{H.-Y.}},
\bauthor{\bsnm{Broughton}, \binits{M.}},
\bauthor{\bsnm{Cotler}, \binits{J.}},
\bauthor{\bsnm{Chen}, \binits{S.}},
\bauthor{\bsnm{Li}, \binits{J.}},
\bauthor{\bsnm{Mohseni}, \binits{M.}},
\bauthor{\bsnm{Neven}, \binits{H.}},
\bauthor{\bsnm{Babbush}, \binits{R.}},
\bauthor{\bsnm{Kueng}, \binits{R.}},
\bauthor{\bsnm{Preskill}, \binits{J.}},
\bauthor{\bsnm{McClean}, \binits{J.R.}}:
\batitle{Quantum advantage in learning from experiments}.
\bjtitle{Science}
\bvolume{376}(\bissue{6598}),
\bfpage{1182}--\blpage{1186}
(\byear{2022})
\doiurl{10.1126/science.abn7293}
\end{barticle}
\endbibitem

\bibitem[\protect\citeauthoryear{Alam et~al.}{2021}]{alam2021}
\begin{botherref}
\oauthor{\bsnm{Alam}, \binits{M.}},
\oauthor{\bsnm{Kundu}, \binits{S.}},
\oauthor{\bsnm{Topaloglu}, \binits{R.O.}},
\oauthor{\bsnm{Ghosh}, \binits{S.}}:
Quantum-{{Classical Hybrid Machine Learning}} for {{Image Classification}}
  ({{ICCAD Special Session Paper}}).
arXiv
(2021).
\doiurl{10.48550/arXiv.2109.02862}
\end{botherref}
\endbibitem

\bibitem[\protect\citeauthoryear{Benedetti et~al.}{2019}]{benedetti2019}
\begin{barticle}
\bauthor{\bsnm{Benedetti}, \binits{M.}},
\bauthor{\bsnm{Lloyd}, \binits{E.}},
\bauthor{\bsnm{Sack}, \binits{S.}},
\bauthor{\bsnm{Fiorentini}, \binits{M.}}:
\batitle{Parameterized quantum circuits as machine learning models}.
\bjtitle{Quantum Science and Technology}
\bvolume{4}(\bissue{4}),
\bfpage{043001}
(\byear{2019})
\doiurl{10.1088/2058-9565/ab4eb5}
\end{barticle}
\endbibitem

\bibitem[\protect\citeauthoryear{Mitarai et~al.}{2018}]{mitarai2018b}
\begin{botherref}
\oauthor{\bsnm{Mitarai}, \binits{K.}},
\oauthor{\bsnm{Negoro}, \binits{M.}},
\oauthor{\bsnm{Kitagawa}, \binits{M.}},
\oauthor{\bsnm{Fujii}, \binits{K.}}:
Quantum circuit learning.
Physical Review A
\textbf{98}(3)
(2018)
\doiurl{10.1103/physreva.98.032309}
\end{botherref}
\endbibitem

\bibitem[\protect\citeauthoryear{Schuld et~al.}{2020}]{schuld2020}
\begin{botherref}
\oauthor{\bsnm{Schuld}, \binits{M.}},
\oauthor{\bsnm{Bocharov}, \binits{A.}},
\oauthor{\bsnm{Svore}, \binits{K.}},
\oauthor{\bsnm{Wiebe}, \binits{N.}}:
Circuit-centric quantum classifiers.
Physical Review A
\textbf{101}(3)
(2020)
\doiurl{10.1103/PhysRevA.101.032308}
{\href{https://arxiv.org/abs/1804.00633}{{arXiv:1804.00633}}}
{[quant-ph]}
\end{botherref}
\endbibitem

\bibitem[\protect\citeauthoryear{McClean et~al.}{2018}]{mcclean2018}
\begin{botherref}
\oauthor{\bsnm{McClean}, \binits{J.R.}},
\oauthor{\bsnm{Boixo}, \binits{S.}},
\oauthor{\bsnm{Smelyanskiy}, \binits{V.N.}},
\oauthor{\bsnm{Babbush}, \binits{R.}},
\oauthor{\bsnm{Neven}, \binits{H.}}:
Barren plateaus in quantum neural network training landscapes.
Nature Communications, Volume 9, Article Number: 4812 (2018)
\textbf{9}(1)
(2018)
\doiurl{10.1038/s41467-018-07090-4}
\end{botherref}
\endbibitem

\bibitem[\protect\citeauthoryear{Sim et~al.}{2019}]{sim2019expressibility}
\begin{barticle}
\bauthor{\bsnm{Sim}, \binits{S.}},
\bauthor{\bsnm{Johnson}, \binits{P.D.}},
\bauthor{\bsnm{Aspuru-Guzik}, \binits{A.}}:
\batitle{Expressibility and entangling capability of parameterized quantum
  circuits for hybrid quantum-classical algorithms}.
\bjtitle{Advanced Quantum Technologies}
\bvolume{2}(\bissue{12}),
\bfpage{1900070}
(\byear{2019})
\end{barticle}
\endbibitem

\bibitem[\protect\citeauthoryear{Hochreiter and
  Schmidhuber}{1997}]{hochreiter1997}
\begin{barticle}
\bauthor{\bsnm{Hochreiter}, \binits{S.}},
\bauthor{\bsnm{Schmidhuber}, \binits{J.}}:
\batitle{Long {{Short-Term Memory}}}.
\bjtitle{Neural Computation}
\bvolume{9}(\bissue{8}),
\bfpage{1735}--\blpage{1780}
(\byear{1997})
\doiurl{10.1162/neco.1997.9.8.1735}
\end{barticle}
\endbibitem

\bibitem[\protect\citeauthoryear{Cho et~al.}{2014}]{cho2014properties}
\begin{botherref}
\oauthor{\bsnm{Cho}, \binits{K.}},
\oauthor{\bsnm{Van~Merri{\"e}nboer}, \binits{B.}},
\oauthor{\bsnm{Bahdanau}, \binits{D.}},
\oauthor{\bsnm{Bengio}, \binits{Y.}}:
On the properties of neural machine translation: Encoder-decoder approaches.
arXiv preprint arXiv:1409.1259
(2014)
\end{botherref}
\endbibitem

\bibitem[\protect\citeauthoryear{Vaswani et~al.}{2017}]{vaswani2017attention}
\begin{botherref}
\oauthor{\bsnm{Vaswani}, \binits{A.}},
\oauthor{\bsnm{Shazeer}, \binits{N.}},
\oauthor{\bsnm{Parmar}, \binits{N.}},
\oauthor{\bsnm{Uszkoreit}, \binits{J.}},
\oauthor{\bsnm{Jones}, \binits{L.}},
\oauthor{\bsnm{Gomez}, \binits{A.N.}},
\oauthor{\bsnm{Kaiser}, \binits{{\L}.}},
\oauthor{\bsnm{Polosukhin}, \binits{I.}}:
Attention is all you need.
Advances in neural information processing systems
\textbf{30}
(2017)
\end{botherref}
\endbibitem

\bibitem[\protect\citeauthoryear{Kitaev et~al.}{2020}]{kitaev2020reformer}
\begin{botherref}
\oauthor{\bsnm{Kitaev}, \binits{N.}},
\oauthor{\bsnm{Kaiser}, \binits{{\L}.}},
\oauthor{\bsnm{Levskaya}, \binits{A.}}:
Reformer: The efficient transformer.
arXiv preprint arXiv:2001.04451
(2020)
\end{botherref}
\endbibitem

\bibitem[\protect\citeauthoryear{Beltagy et~al.}{2020}]{beltagy2020longformer}
\begin{botherref}
\oauthor{\bsnm{Beltagy}, \binits{I.}},
\oauthor{\bsnm{Peters}, \binits{M.E.}},
\oauthor{\bsnm{Cohan}, \binits{A.}}:
Longformer: The long-document transformer.
arXiv preprint arXiv:2004.05150
(2020)
\end{botherref}
\endbibitem

\bibitem[\protect\citeauthoryear{Wang et~al.}{2020}]{wang2020linformer}
\begin{botherref}
\oauthor{\bsnm{Wang}, \binits{S.}},
\oauthor{\bsnm{Li}, \binits{B.Z.}},
\oauthor{\bsnm{Khabsa}, \binits{M.}},
\oauthor{\bsnm{Fang}, \binits{H.}},
\oauthor{\bsnm{Ma}, \binits{H.}}:
Linformer: Self-attention with linear complexity.
arXiv preprint arXiv:2006.04768
(2020)
\end{botherref}
\endbibitem

\bibitem[\protect\citeauthoryear{Choromanski
  et~al.}{2020}]{choromanski2020rethinking}
\begin{botherref}
\oauthor{\bsnm{Choromanski}, \binits{K.}},
\oauthor{\bsnm{Likhosherstov}, \binits{V.}},
\oauthor{\bsnm{Dohan}, \binits{D.}},
\oauthor{\bsnm{Song}, \binits{X.}},
\oauthor{\bsnm{Gane}, \binits{A.}},
\oauthor{\bsnm{Sarlos}, \binits{T.}},
\oauthor{\bsnm{Hawkins}, \binits{P.}},
\oauthor{\bsnm{Davis}, \binits{J.}},
\oauthor{\bsnm{Mohiuddin}, \binits{A.}},
\oauthor{\bsnm{Kaiser}, \binits{L.}}, et al.:
Rethinking attention with performers.
arXiv preprint arXiv:2009.14794
(2020)
\end{botherref}
\endbibitem

\bibitem[\protect\citeauthoryear{Gu et~al.}{2021}]{gu2021efficiently}
\begin{botherref}
\oauthor{\bsnm{Gu}, \binits{A.}},
\oauthor{\bsnm{Goel}, \binits{K.}},
\oauthor{\bsnm{R{\'e}}, \binits{C.}}:
Efficiently modeling long sequences with structured state spaces.
arXiv preprint arXiv:2111.00396
(2021)
\end{botherref}
\endbibitem

\bibitem[\protect\citeauthoryear{Smith et~al.}{2022}]{smith2022simplified}
\begin{botherref}
\oauthor{\bsnm{Smith}, \binits{J.T.}},
\oauthor{\bsnm{Warrington}, \binits{A.}},
\oauthor{\bsnm{Linderman}, \binits{S.W.}}:
Simplified state space layers for sequence modeling.
arXiv preprint arXiv:2208.04933
(2022)
\end{botherref}
\endbibitem

\bibitem[\protect\citeauthoryear{Gu and Dao}{2023}]{gu2023mamba}
\begin{botherref}
\oauthor{\bsnm{Gu}, \binits{A.}},
\oauthor{\bsnm{Dao}, \binits{T.}}:
Mamba: Linear-time sequence modeling with selective state spaces.
arXiv preprint arXiv:2312.00752
(2023)
\end{botherref}
\endbibitem

\bibitem[\protect\citeauthoryear{Chen et~al.}{2020}]{chen2020b}
\begin{botherref}
\oauthor{\bsnm{Chen}, \binits{S.Y.-C.}},
\oauthor{\bsnm{Yoo}, \binits{S.}},
\oauthor{\bsnm{Fang}, \binits{Y.-L.L.}}:
Quantum Long Short-Term Memory.
arXiv
(2020).
\doiurl{10.48550/arXiv.2009.01783}
\end{botherref}
\endbibitem

\bibitem[\protect\citeauthoryear{Li et~al.}{2023}]{li2023}
\begin{botherref}
\oauthor{\bsnm{Li}, \binits{Y.}},
\oauthor{\bsnm{Wang}, \binits{Z.}},
\oauthor{\bsnm{Han}, \binits{R.}},
\oauthor{\bsnm{Shi}, \binits{S.}},
\oauthor{\bsnm{Li}, \binits{J.}},
\oauthor{\bsnm{Shang}, \binits{R.}},
\oauthor{\bsnm{Zheng}, \binits{H.}},
\oauthor{\bsnm{Zhong}, \binits{G.}},
\oauthor{\bsnm{Gu}, \binits{Y.}}:
Quantum {{Recurrent Neural Networks}} for {{Sequential Learning}}.
arXiv
(2023).
\doiurl{10.48550/arXiv.2302.03244}
\end{botherref}
\endbibitem

\bibitem[\protect\citeauthoryear{Moon et~al.}{2025}]{moon2025}
\begin{botherref}
\oauthor{\bsnm{Moon}, \binits{K.-H.}},
\oauthor{\bsnm{Jeong}, \binits{S.-G.}},
\oauthor{\bsnm{Hwang}, \binits{W.-J.}}:
{{QSegRNN}}: Quantum segment recurrent neural network for time series
  forecasting.
EPJ Quantum Technology
\textbf{12}(1)
(2025)
\doiurl{10.1140/epjqt/s40507-025-00333-6}
\end{botherref}
\endbibitem

\bibitem[\protect\citeauthoryear{Basile and Tamburini}{2017}]{basile2017}
\begin{bchapter}
\bauthor{\bsnm{Basile}, \binits{I.}},
\bauthor{\bsnm{Tamburini}, \binits{F.}}:
\bctitle{Towards quantum language models}.
In: \bbtitle{Proceedings of the 2017 {{Conference}} on {{Empirical Methods}} in
  {{Natural}} {{Language Processing}}},
pp. \bfpage{1840}--\blpage{1849}.
\bpublisher{Association for Computational Linguistics},
\blocation{Copenhagen, Denmark}
(\byear{2017}).
\doiurl{10.18653/v1/D17-1196}
\end{bchapter}
\endbibitem

\bibitem[\protect\citeauthoryear{Sipio et~al.}{2021}]{sipio2021}
\begin{botherref}
\oauthor{\bsnm{Sipio}, \binits{R.D.}},
\oauthor{\bsnm{Huang}, \binits{J.-H.}},
\oauthor{\bsnm{Chen}, \binits{S.Y.-C.}},
\oauthor{\bsnm{Mangini}, \binits{S.}},
\oauthor{\bsnm{Worring}, \binits{M.}}:
The {{Dawn}} of {{Quantum Natural Language Processing}}.
arXiv
(2021).
\doiurl{10.48550/arXiv.2110.06510}
\end{botherref}
\endbibitem

\bibitem[\protect\citeauthoryear{Emmanoulopoulos and
  Dimoska}{2022}]{emmanoulopoulos2022c}
\begin{botherref}
\oauthor{\bsnm{Emmanoulopoulos}, \binits{D.}},
\oauthor{\bsnm{Dimoska}, \binits{S.}}:
Quantum {{Machine Learning}} in {{Finance}}: {{Time Series Forecasting}}.
arXiv
(2022).
\doiurl{10.48550/arXiv.2202.00599}
\end{botherref}
\endbibitem

\bibitem[\protect\citeauthoryear{Comajoan~Cara et~al.}{2024}]{comajoancara2024}
\begin{barticle}
\bauthor{\bsnm{Comajoan~Cara}, \binits{M.}},
\bauthor{\bsnm{Dahale}, \binits{G.R.}},
\bauthor{\bsnm{Dong}, \binits{Z.}},
\bauthor{\bsnm{Forestano}, \binits{R.T.}},
\bauthor{\bsnm{Gleyzer}, \binits{S.}},
\bauthor{\bsnm{Justice}, \binits{D.}},
\bauthor{\bsnm{Kong}, \binits{K.}},
\bauthor{\bsnm{Magorsch}, \binits{T.}},
\bauthor{\bsnm{Matchev}, \binits{K.T.}},
\bauthor{\bsnm{Matcheva}, \binits{K.}},
\bauthor{\bsnm{Unlu}, \binits{E.B.}}:
\batitle{Quantum {{Vision Transformers}} for {{Quark}}--{{Gluon
  Classification}}}.
\bjtitle{Axioms}
\bvolume{13}(\bissue{5}),
\bfpage{323}
(\byear{2024})
\doiurl{10.3390/axioms13050323}
\end{barticle}
\endbibitem

\bibitem[\protect\citeauthoryear{Guo et~al.}{2024}]{guo2024quantum}
\begin{botherref}
\oauthor{\bsnm{Guo}, \binits{N.}},
\oauthor{\bsnm{Yu}, \binits{Z.}},
\oauthor{\bsnm{Choi}, \binits{M.}},
\oauthor{\bsnm{Agrawal}, \binits{A.}},
\oauthor{\bsnm{Nakaji}, \binits{K.}},
\oauthor{\bsnm{Aspuru-Guzik}, \binits{A.}},
\oauthor{\bsnm{Rebentrost}, \binits{P.}}:
Quantum linear algebra is all you need for transformer architectures.
arXiv preprint arXiv:2402.16714
(2024)
\end{botherref}
\endbibitem

\bibitem[\protect\citeauthoryear{Khatri et~al.}{2024}]{khatri2024}
\begin{botherref}
\oauthor{\bsnm{Khatri}, \binits{N.}},
\oauthor{\bsnm{Matos}, \binits{G.}},
\oauthor{\bsnm{Coopmans}, \binits{L.}},
\oauthor{\bsnm{Clark}, \binits{S.}}:
Quixer: {{A Quantum Transformer Model}}.
arXiv
(2024).
\doiurl{10.48550/arXiv.2406.04305}
\end{botherref}
\endbibitem

\bibitem[\protect\citeauthoryear{Gu et~al.}{2020}]{Gu2020}
\begin{botherref}
\oauthor{\bsnm{Gu}, \binits{A.}},
\oauthor{\bsnm{Dao}, \binits{T.}},
\oauthor{\bsnm{Ermon}, \binits{S.}},
\oauthor{\bsnm{Rudra}, \binits{A.}},
\oauthor{\bsnm{Re}, \binits{C.}}:
Hippo: Recurrent memory with optimal polynomial projections
(2020)
\doiurl{10.48550/ARXIV.2008.07669}
{\href{https://arxiv.org/abs/2008.07669}{{arXiv:2008.07669}}}
{[cs.LG]}
\end{botherref}
\endbibitem

\bibitem[\protect\citeauthoryear{Gu et~al.}{2022}]{Gu2022}
\begin{botherref}
\oauthor{\bsnm{Gu}, \binits{A.}},
\oauthor{\bsnm{Gupta}, \binits{A.}},
\oauthor{\bsnm{Goel}, \binits{K.}},
\oauthor{\bsnm{Ré}, \binits{C.}}:
On the parameterization and initialization of diagonal state space models
(2022)
\doiurl{10.48550/ARXIV.2206.11893}
{\href{https://arxiv.org/abs/2206.11893}{{arXiv:2206.11893}}}
{[cs.LG]}
\end{botherref}
\endbibitem

\bibitem[\protect\citeauthoryear{Bergholm et~al.}{2022}]{bergholm2022}
\begin{botherref}
\oauthor{\bsnm{Bergholm}, \binits{V.}},
\oauthor{\bsnm{Izaac}, \binits{J.}},
\oauthor{\bsnm{Schuld}, \binits{M.}},
\oauthor{\bsnm{Gogolin}, \binits{C.}},
\oauthor{\bsnm{Ahmed}, \binits{S.}},
\oauthor{\bsnm{Ajith}, \binits{V.}},
\oauthor{\bsnm{Alam}, \binits{M.S.}},
\oauthor{\bsnm{{Alonso-Linaje}}, \binits{G.}},
\oauthor{\bsnm{AkashNarayanan}, \binits{B.}},
\oauthor{\bsnm{Asadi}, \binits{A.}},
\oauthor{\bsnm{Arrazola}, \binits{J.M.}},
\oauthor{\bsnm{Azad}, \binits{U.}},
\oauthor{\bsnm{Banning}, \binits{S.}},
\oauthor{\bsnm{Blank}, \binits{C.}},
\oauthor{\bsnm{Bromley}, \binits{T.R.}},
\oauthor{\bsnm{Cordier}, \binits{B.A.}},
\oauthor{\bsnm{Ceroni}, \binits{J.}},
\oauthor{\bsnm{Delgado}, \binits{A.}},
\oauthor{\bsnm{Matteo}, \binits{O.D.}},
\oauthor{\bsnm{Dusko}, \binits{A.}},
\oauthor{\bsnm{Garg}, \binits{T.}},
\oauthor{\bsnm{Guala}, \binits{D.}},
\oauthor{\bsnm{Hayes}, \binits{A.}},
\oauthor{\bsnm{Hill}, \binits{R.}},
\oauthor{\bsnm{Ijaz}, \binits{A.}},
\oauthor{\bsnm{Isacsson}, \binits{T.}},
\oauthor{\bsnm{Ittah}, \binits{D.}},
\oauthor{\bsnm{Jahangiri}, \binits{S.}},
\oauthor{\bsnm{Jain}, \binits{P.}},
\oauthor{\bsnm{Jiang}, \binits{E.}},
\oauthor{\bsnm{Khandelwal}, \binits{A.}},
\oauthor{\bsnm{Kottmann}, \binits{K.}},
\oauthor{\bsnm{Lang}, \binits{R.A.}},
\oauthor{\bsnm{Lee}, \binits{C.}},
\oauthor{\bsnm{Loke}, \binits{T.}},
\oauthor{\bsnm{Lowe}, \binits{A.}},
\oauthor{\bsnm{McKiernan}, \binits{K.}},
\oauthor{\bsnm{Meyer}, \binits{J.J.}},
\oauthor{\bsnm{{Monta{\~n}ez-Barrera}}, \binits{J.A.}},
\oauthor{\bsnm{Moyard}, \binits{R.}},
\oauthor{\bsnm{Niu}, \binits{Z.}},
\oauthor{\bsnm{O'Riordan}, \binits{L.J.}},
\oauthor{\bsnm{Oud}, \binits{S.}},
\oauthor{\bsnm{Panigrahi}, \binits{A.}},
\oauthor{\bsnm{Park}, \binits{C.-Y.}},
\oauthor{\bsnm{Polatajko}, \binits{D.}},
\oauthor{\bsnm{Quesada}, \binits{N.}},
\oauthor{\bsnm{Roberts}, \binits{C.}},
\oauthor{\bsnm{S{\'a}}, \binits{N.}},
\oauthor{\bsnm{Schoch}, \binits{I.}},
\oauthor{\bsnm{Shi}, \binits{B.}},
\oauthor{\bsnm{Shu}, \binits{S.}},
\oauthor{\bsnm{Sim}, \binits{S.}},
\oauthor{\bsnm{Singh}, \binits{A.}},
\oauthor{\bsnm{Strandberg}, \binits{I.}},
\oauthor{\bsnm{Soni}, \binits{J.}},
\oauthor{\bsnm{Sz{\'a}va}, \binits{A.}},
\oauthor{\bsnm{Thabet}, \binits{S.}},
\oauthor{\bsnm{{Vargas-Hern{\'a}ndez}}, \binits{R.A.}},
\oauthor{\bsnm{Vincent}, \binits{T.}},
\oauthor{\bsnm{Vitucci}, \binits{N.}},
\oauthor{\bsnm{Weber}, \binits{M.}},
\oauthor{\bsnm{Wierichs}, \binits{D.}},
\oauthor{\bsnm{Wiersema}, \binits{R.}},
\oauthor{\bsnm{Willmann}, \binits{M.}},
\oauthor{\bsnm{Wong}, \binits{V.}},
\oauthor{\bsnm{Zhang}, \binits{S.}},
\oauthor{\bsnm{Killoran}, \binits{N.}}:
{{PennyLane}}: {{Automatic}} Differentiation of Hybrid Quantum-Classical
  Computations.
arXiv
(2022).
\doiurl{10.48550/arXiv.1811.04968}
\end{botherref}
\endbibitem

\bibitem[\protect\citeauthoryear{Schuld et~al.}{2020}]{schuld2020circuit}
\begin{barticle}
\bauthor{\bsnm{Schuld}, \binits{M.}},
\bauthor{\bsnm{Bocharov}, \binits{A.}},
\bauthor{\bsnm{Svore}, \binits{K.M.}},
\bauthor{\bsnm{Wiebe}, \binits{N.}}:
\batitle{Circuit-centric quantum classifiers}.
\bjtitle{Physical Review A}
\bvolume{101}(\bissue{3}),
\bfpage{032308}
(\byear{2020})
\end{barticle}
\endbibitem

\bibitem[\protect\citeauthoryear{Paszke et~al.}{2019}]{paszke2019PyTorch}
\begin{botherref}
\oauthor{\bsnm{Paszke}, \binits{A.}},
\oauthor{\bsnm{Gross}, \binits{S.}},
\oauthor{\bsnm{Massa}, \binits{F.}},
\oauthor{\bsnm{Lerer}, \binits{A.}},
\oauthor{\bsnm{Bradbury}, \binits{J.}},
\oauthor{\bsnm{Chanan}, \binits{G.}},
\oauthor{\bsnm{Killeen}, \binits{T.}},
\oauthor{\bsnm{Lin}, \binits{Z.}},
\oauthor{\bsnm{Gimelshein}, \binits{N.}},
\oauthor{\bsnm{Antiga}, \binits{L.}},
\oauthor{\bsnm{Desmaison}, \binits{A.}},
\oauthor{\bsnm{K{\"o}pf}, \binits{A.}},
\oauthor{\bsnm{Yang}, \binits{E.}},
\oauthor{\bsnm{DeVito}, \binits{Z.}},
\oauthor{\bsnm{Raison}, \binits{M.}},
\oauthor{\bsnm{Tejani}, \binits{A.}},
\oauthor{\bsnm{Chilamkurthy}, \binits{S.}},
\oauthor{\bsnm{Steiner}, \binits{B.}},
\oauthor{\bsnm{Fang}, \binits{L.}},
\oauthor{\bsnm{Bai}, \binits{J.}},
\oauthor{\bsnm{Chintala}, \binits{S.}}:
{{PyTorch}}: {{An Imperative Style}}, {{High-Performance Deep Learning
  Library}}.
arXiv
(2019).
\doiurl{10.48550/arXiv.1912.01703}
\end{botherref}
\endbibitem

\end{thebibliography}

\end{document}